\documentclass[preprint,review,accept,oneauthor,pdftex,galaxies]{mdpi}

\firstpage{1} 
\pubvolume{xx}
\issuenum{1}
\articlenumber{5}
\pubyear{2019}
\copyrightyear{2019}
\history{Received: date; Accepted: date; Published: date}

\continuouspages{yes}

\usepackage{color}

\Title{A Census of B[e] Supergiants}

\Author{Michaela Kraus $^{1}$\orcidA{}}

\AuthorNames{Michaela Kraus}

\address{%
$^{1}$ \quad Astronomical Institute, Czech Academy of Sciences, 
 Fri\v{c}ova 298, 251\,65 Ond\v{r}ejov, Czech Republic; michaela.kraus@asu.cas.cz}

\corres{Correspondence: michaela.kraus@asu.cas.cz}

\abstract{Stellar evolution theory is most uncertain for massive stars. For reliable predictions of the evolution of massive stars and their final fate, solid constraints on the physical parameters, and their changes along the evolution and in different environments, are required. Massive stars evolve through a variety of short transition phases, in which they can experience large mass-loss either in the form of dense winds or via sudden eruptions. The B[e] supergiants comprise one such group of massive transition objects. They are characterized by dense, dusty disks of yet unknown origin. In the Milky Way, identification and classification of B[e] supergiants is usually hampered by their uncertain distances hence luminosities, and by the confusion of low-luminosity candidates with massive pre-main sequence objects. The extragalactic objects are often mistaken as  quiescent or candidate luminous blue variables, with whom B[e] supergiants share a number of spectroscopic characteristics. In this review, proper criteria are provided, based on which B[e] supergiants can be unambiguously classified and separated from  other high luminosity post-main sequence stars and pre-main sequence stars. Using these criteria, the B[e] supergiant samples in diverse galaxies are critically inspected, to achieve a reliable census of the current population.}


\begin{document}



\section{Introduction}
Massive stars play a major role in the evolution of their host galaxies. 
Via stellar winds, they strongly enrich the interstellar medium (ISM) with
chemically processed material and deposit large amounts of momentum and energy into their
surroundings during their entire lifetime, from the main-sequence up to their final fate as
spectacular supernova explosions (e.g. \cite{2013A&A...558L...1G, 2013A&A...558A.131G, 
2013A&A...550L...7G}). The released energy provides the 
ionizing radiation, substantially supplies the global energy budget of the host galaxy and 
significantly contributes to shaping the local ISM, whereas the released material condenses into 
molecules and dust, providing the cradles for the next generation of stars and planets (e.g., 
\cite{2012A&A...538A..75P, 2014MNRAS.445..581H}).

Despite their great importance, stellar evolution theory is most uncertain for massive stars due to 
the often still poor understanding of some physical processes in the stellar interiors (e.g., core 
convective overshooting, chemical diffusion, internal differential rotation law and angular momentum 
transport), the excitation and propagation of pulsation instabilities within their atmospheres, the 
amount of mass loss via stellar (often asymmetric) winds and (irregular) mass 
ejections, and the role of binarity for certain phases. 

From an observational point of view, the post-main sequence domain within the Hertzsprung-Russell (HR) 
diagram is populated with various types of extreme massive stars. These are found to be in  
transition phases, in which the stars shed huge amounts of material into their environments, typically 
via episodic, sometimes even eruptive events. These objects are luminous super- or hypergiants 
populating the upper part of the HR diagram and spreading from spectral type O to F or even later.
The ejected material thereby accumulates in either nebulae, shells, or even disk-like structures. 

The mass-loss of massive stars not only critically depends on the physical parameters, such as 
mass, effective temperature, and rotation speed, but also on the chemical composition of the star.
The amount of mass that is lost, within each individual evolutionary phase, determines the fate of the 
object. It is thus not surprising that relative numbers of various types of massive stars can change 
drastically among galaxies with different metallicities (e.g. \cite{2016AJ....152...62M}). For reliable 
predictions of the evolutionary path of massive stars in any environment, solid constraints on the 
physical parameters, used in modern stellar evolution models, are indispensable. To obtain such 
constraints, the properties of the members within each class of objects need to be studied in great 
detail and within a variety of environments. This requires statistically significant 
samples of stars in each class of objects, suitable for a detailed analysis. The star-forming galaxies
within the local Universe, in which the metallicities spread over a factor of about 25 between the 
most metal poor and the most metal rich representative, are the most ideal sites to tackle this 
challenge. 

This review is devoted to the B[e] supergiants which comprise one of the various classes of extreme 
massive stars in transition. The article is structured as follows: First an overview on the general 
properties of these stars is given based mostly on the well-studied sample within the Magellanic 
Clouds (Section \ref{sect:B[e]SGs}), followed by a review on how these objects are searched for in 
various environments (Section \ref{sect:method}). A census of the currently known objects in the Local 
Group galaxies and slightly beyond is presented in Section \ref{sect:census}, based on a critical 
inspection of the properties of the individual candidates. The discussion of the B[e]SG samples 
and our conclusions are finally summarized in Section \ref{sect:discussion}.

\begin{figure}[H]
\centering
\includegraphics[width=11cm]{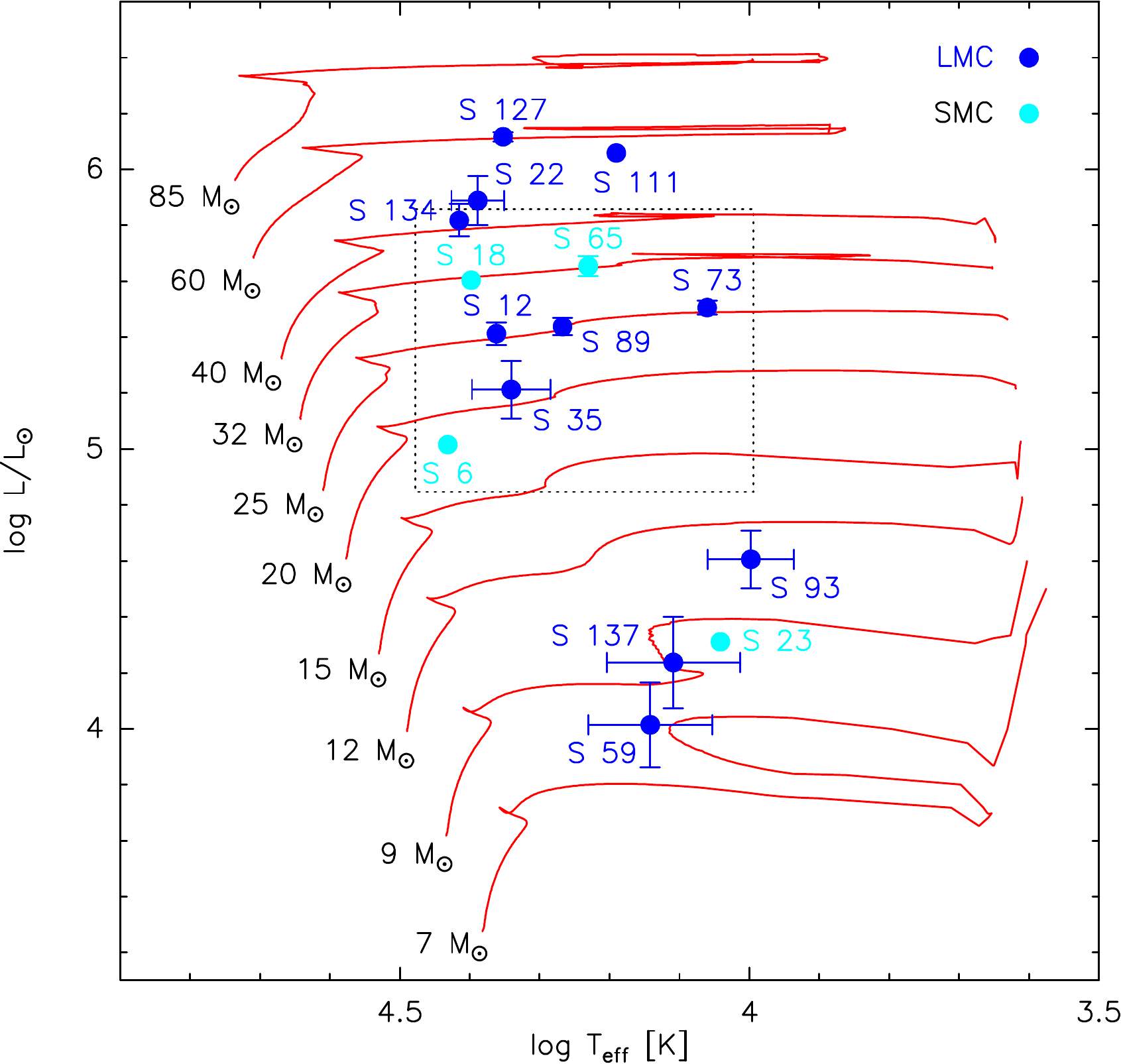}
\caption{HR diagram showing the positions of the classical MC B[e]SG sample \cite{2006ASPC..355..135Z}. 
Also included are the stellar evolutionary tracks at SMC metallicity for stars rotating initially with 
40\% of their critical velocity (from \cite{2013A&A...558A.103G}). The dotted square contains objects 
that display CO band emission (except for S\,89, see Section\,\ref{sect:current_evol}). For brevity and 
readability, the identifiers LHA 120 and LHA 115 for objects within the LMC and SMC, respectively, 
have been omitted.}
\label{fig:HRD-classic}
\end{figure}

\section{B[e] supergiants}\label{sect:B[e]SGs}

The early-type supergiants include a class of emission-line objects, whose optical spectra 
display a peculiar character with strong Balmer emission along with narrow emission lines from 
permitted and forbidden transitions (e.g. \cite{1924BHarO.801....1C, 1956ApJS....2..315H, 
1957PASP...69..137S, 1960MNRAS.121..337F}). 
The latter are indicative of a cool and slowly expanding medium. With the advent of ultraviolet (UV) 
observations taken with the International Ultraviolet Explorer (IUE), these stars were found to 
display very broad blueshifted resonance lines of highly ionized elements in this spectral range. 
These resonance lines originate from a hot and fast stellar line-driven wind which is very typical for 
supergiants in this temperature and luminosity range. 

Another peculiar property of these stars was discovered in the near-infrared, in which these 
objects possess a pronounced excess emission pointing to hot circumstellar dust 
(\cite{1976ApJ...210..666A, 1977MNRAS.178P...9G, 1983A&A...120..287S, 1984A&A...131L...5S, 
1985A&A...143..421Z, 1986A&A...163..119Z}). This dust was proposed to be 
most likely produced within the slow and cool component and to possibly populate a ring or 
disk-like region at far distances from the luminous central objects (\cite{1985A&A...143..421Z}).

\begin{table}[H]
\caption{Names and parameters of the established B[e]SG sample within the Magellanic Clouds.}
\centering
\begin{tabular}{llccccc}
\toprule
\textbf{Object}	& \textbf{Other Common Identifiers} & \textbf{log} \sl{\textbf{T$_{\rm\bf eff}$}} & \textbf{log} \sl{\textbf{L/L$_{\bf{\odot}}$}} & \sl{\textbf{V}} & \sl{\textbf{E(B-V)}} & \textbf{Ref.} \\
\midrule
\multicolumn{7}{c}{Large Magellanic Cloud B[e]SGs}\\
\midrule
LHA\,120-S\,12  & SK $-$67 23              & 4.36          &  5.41$\pm$0.04 & 12.6 & 0.2--0.25  & \protect{\cite{1986A&A...163..119Z}} \\
LHA\,120-S\,22  & HD\,34664, SK $-$67 64   & 4.39$\pm$0.04 &  5.89$\pm$0.09 & 11.7 & 0.25--0.3  & \protect{\cite{1986A&A...163..119Z}} \\
LHA\,120-S\,35  & SK $-$66 97              & 4.34$\pm$0.06 &  5.21$\pm$0.10 & 12.5 & 0.07  & \protect{\cite{1995A&A...302..409G}}  \\
LHA\,120-S\,59  & \ldots                   & 4.14$\pm$0.09 &  4.01$\pm$0.15 & 14.4 & 0.05  & \protect{\cite{1995A&A...302..409G}}  \\
LHA\,120-S\,73  & RMC\,66, HD\,268835      & 4.06          &  5.51$\pm$0.03 & 10.6 & 0.12--0.15  & \protect{\cite{1983A&A...120..287S}} \\
LHA\,120-S\,89  & RMC\,82, HD\,269217      & 4.27          &  5.44$\pm$0.03 & 12.0 & 0.20  & \protect{\cite{1986A&A...163..119Z}}\\
LHA\,120-S\,93  & SK $-$68 66              & 4.00$\pm$0.06 &  4.61$\pm$0.10 & 12.7 & 0.22  & \protect{\cite{1995A&A...302..409G}} \\
LHA\,120-S\,111$^{a}$ & HD\,269599         & 4.19          &  6.06          & 10.3 & 0.28  & \protect{\cite{1988ApJ...334..639M}}  \\
LHA\,120-S\,127 & RMC\,126, HD\,37974      & 4.35          &  6.12$\pm$0.02 & 10.9 & 0.25  & \protect{\cite{1985A&A...143..421Z}}  \\
LHA\,120-S\,134$^{b,c}$ & HD\,38489, SK $-$69 259  & 4.42          &  5.82$\pm$0.06 & 12.0 & 0.2--0.25  & \protect{\cite{1986A&A...163..119Z}} \\
LHA\,120-S\,137 & \ldots                   & 4.11$\pm$0.10 &  4.24$\pm$0.16 & 14.0 & 0.17  & \protect{\cite{1995A&A...302..409G}}  \\
\midrule
\multicolumn{7}{c}{Small Magellanic Cloud B[e]SGs}\\
\midrule
LHA\,115-S\,6$^{b}$   & RMC\,4, AzV\,16          & 4.43          &  5.02          & 13.0 & 0.07 & \protect{\cite{1996A&A...309..505Z}} \\
LHA\,115-S\,18$^{b,c}$  & AzV\,154                 & 4.40          &  5.60          & 13.3 & 0.4 & \protect{\cite{1989A&A...220..206Z}} \\
LHA\,115-S\,23  & AzV\,172                 & 4.04          &  4.31          & 13.3 & 0.03--0.1  & \protect{\cite{1992A&A...260..205Z}}  \\
LHA\,115-S\,65  & RMC\,50                  & 4.23          &  5.65$\pm$0.04 & 11.6 & 0.15--0.2  & \protect{\cite{1986A&A...163..119Z}} \\
\bottomrule
\multicolumn{7}{p{.9\textwidth}}{{\bf Note:} Former designations of some of the objects as Hen\,S\,\# (see, e.g., \cite{Lamers1998}) were omitted here, as these are not SIMBAD identifiers. Still, the LHA\,120-S respectively LHA\,115-S and the former Hen\,S numbers refer to the same objects.}\\
\multicolumn{7}{p{.9\textwidth}}{$^{a}$ The star is also listed as RMC\,105 in SIMBAD, but this 
designation should be used for a neighboring, normal B-type star in this dense cluster (see 
\cite{1984Msngr..38...28A}).}\\
\multicolumn{7}{p{.9\textwidth}}{{$^{b}$}Confirmed or suspected binary.}\\
\multicolumn{7}{p{.9\textwidth}}{{$^{c}$}X-ray source 
(\protect{\cite{2013A&A...560A..10C, 2014ApJ...788...83M, 2015salt.confE..55B}}).}\\
\label{tab:logTL}
\end{tabular}
\end{table}

In the HR diagram these objects are all found beyond the main-sequence and with luminosities 
spreading from about $\log L/L_{\odot} \sim$ 4 to about $\log L/L_{\odot} \sim$ 6 implying that they
are all evolved, massive stars. This luminosity range was determined from the sample residing in the
Magellanic Clouds (MCs), for which the luminosity determination is unquestionable, due to the low 
extinction towards the MCs and their well constraint distances. The classification of Galactic objects 
as supergiants bears much higher uncertainties due to their often poorly constraint distances hence 
luminosities. We will come back to this issue in Section\,\ref{sect:MilkyWay}.

The position of the MC sample in the HR diagram is shown in Figure\,\ref{fig:HRD-classic} for the 
values of luminosity and effective temperature listed in Table\,\ref{tab:logTL}. The stellar 
parameters (effective temperature $T_{\rm eff}$, visual magnitude $V$, and color excess $E(B-V)$) of 
the sample have been taken from the references listed in the last column of Table\,\ref{tab:logTL}. 
For the calculations of the luminosities, distance moduli of 18.5 and 18.9\,mag for the Large 
respectively Small Magellanic Clouds have been utilized (see the review paper by Humphreys, this 
volume) along with bolometric corrections from \cite{2005oasp.book.....G}. 

The presence of dust around an early-type (typically of spectral type B) supergiant, along with the 
often pure emission-line spectra with numerous forbidden lines predominantly of [Fe\,{\sc ii}] and 
[O\,{\sc i}] finally resulted in the designation of these objects as B[e] supergiants 
(B[e]SGs)\footnote{Note that these objects have previously been abbreviated sgB[e] (\cite{Lamers1998}), 
to separate them from other stars showing the B[e] phenomenon. We prefer the designation B[e]SG, to be 
in line with the naming and abbreviation of other type of supergiants such as blue supergiant (BSG) and 
red supergiant (RSG).}.

\subsection{General aspects of B[e]SG stars' disks}
 
There are compelling evidences that B[e]SGs are surrounded by gaseous and dusty disks. 
The simultaneous presence of a hot and fast polar wind traced in the UV, and the cool and slow 
equatorial wind traced at optical wavelengths led to the assignment of a so-called hybrid or 
two-component wind model (\cite{1985A&A...143..421Z}). For this two-component wind a density contrast 
between the equatorial and polar components of 100--1000 was proposed, meaning that the equatorial 
wind might be assigned the character of an outflowing disk (\cite{1989A&A...220..206Z}).

The degree of non-sphericity of the envelopes, respectively the latitude dependence of the wind 
density is pursued by the measured net intrinsic polarization  (\cite{1992ApJ...398..286M, 
2001A&A...377..581M}) and from spectropolarimetric observations (\cite{2006ASPC..355..147M, 
2017ASPC..508..109S}). The often high degree of intrinsic polarization support the idea of a combination 
of Thomson scattering by free electrons and Mie scattering by dust in a circumstellar disk 
(\cite{1989A&A...214..274Z}). 

If the disks of B[e]SGs are supposed to form from a high-density equatorial stellar outflow, there 
should be a transition zone between the atomic gas and the location of the dust, in which molecules 
can form in substantial amounts, because the high gas density can shield the material from the direct 
irradiation with dissociating UV photons coming from the hot luminous star. And in fact, molecular 
emission, in particular of the first-overtone bands of carbon monoxide (CO), has been detected in the 
$K$-band spectra of a number of B[e]SGs in the Galaxy and the MCs (e.g., \cite{1988ApJ...334..639M, 
1988ApJ...324.1071M, 1989A&A...223..237M, 1996ApJ...470..597M, 2012A&A...543A..77W, 2012BAAA...55..123M,
2012MNRAS.426L..56O, 2013A&A...558A..17O, 2013A&A...549A..28K, 2016A&A...593A.112K, 
2018A&A...612A.113T}, see Table\,\ref{tab:CO-O-Ca}). To produce the 
characteristic observed emission spectra with several individual band heads, temperatures of the CO 
gas higher than $\sim$ 2000\,K are required. These temperatures are in excess of the dust sublimation 
temperature, which is on the order of $\sim$ 1500\,K, placing the CO emitting region closer to the 
star than the dust.

Additional hot molecular emission from silicon oxide (SiO) has been identified in four Galactic 
B[e]SGs (\cite{2015ApJ...800L..20K}), and a feature arising in the optical spectrum, that has been 
tentatively identified as emission from titanium oxide (TiO), was reported from five MC B[e]SGs 
(\cite{1989A&A...220..206Z, 2012MNRAS.427L..80T, 2018A&A...612A.113T, 2016A&A...593A.112K}). However, to 
date no systematic surveys for molecular emission has been performed, so that these numbers are not 
representative for the existence or absence of molecules in the environments of B[e]SGs. For instance, 
SiO emission has not been searched for yet in any of the MC B[e]SGs, and only those Galactic B[e]SGs 
with the most intense CO band emission have been observed in the wavelength range of the first-overtone 
band of SiO arising in the $L$-band.  Hence, one might expect to find molecular emission from SiO in many 
more objects, but also emission from other yet undiscovered molecules that might form in the 
environments of B[e]SGs. What is interesting though is the fact that all MC stars displaying TiO 
emission also have CO emission, whereas the opposite does not hold. No detection of TiO from 
Galactic B[e]SGs has been reported so far. 

Finally, the power of optical interferometry operating at near- and mid-infrared wavelengths should 
be mentioned when talking about the disks of B[e]SGs. Based on this technique, the disks of the 
closest and infrared brightest Galactic objects could be spatially resolved, providing precise 
measurements of the disk inclinations, disk sizes, and the distances of the emitting material 
(dust, CO gas, and ionized gas traced by the Br\,$\gamma$ emission) from the central star (see 
\cite{2007A&A...464...81D, 2011A&A...525A..22D, 2011A&A...526A.107M, 2012A&A...548A..72C, 
2012A&A...545L..10W, 2012A&A...543A..77W, 2012A&A...538A...6W}).

\begin{table}[H]
\caption{Presence of disk tracers in the optical and near-IR spectra of the Galactic and Magellanic 
Cloud B[e]SG samples.}
\centering
\begin{tabular}{lccccccccc}
\toprule
\textbf{Object} &\textbf{[Ca\,{\sc ii}]} & \textbf{[O\,{\sc i}]$^{a}$} & \textbf{Ref.} & \textbf{CO} & \textbf{$^{12}$C/$^{13}$C}  & \textbf{Ref.} & \textbf{TiO} & \textbf{SiO} & \textbf{Ref.} \\
\midrule
\multicolumn{10}{c}{Large Magellanic Cloud B[e]SGs}\\
\midrule
LHA\,120-S\,12 &  yes  & no  & \protect{\cite{2012MNRAS.423..284A}}    &  yes     & 20$\pm$2   & \protect{\cite{1988ApJ...334..639M, 2013A&A...558A..17O}} & yes & \ldots & \protect{\cite{1989A&A...220..206Z}} \\
LHA\,120-S\,22 &  yes  & yes & \protect{\cite{2012MNRAS.423..284A}}    &  no      & \ldots     & \protect{\cite{1988ApJ...334..639M, 2013A&A...558A..17O}} & no & \ldots & TW$^{b}$  \\
LHA\,120-S\,35 &  yes  & yes & \protect{\cite{2018A&A...612A.113T}}  &  yes     & 10$\pm$2   & \protect{\cite{2013A&A...558A..17O, 2018A&A...612A.113T}} & yes & \ldots & \protect{\cite{2018A&A...612A.113T}} \\
LHA\,120-S\,59 &  no   & yes & \protect{\cite{2019MNRAS.488.1090C}} &  ?       & \ldots     & \protect{\cite{2013A&A...558A..17O}} & no & \ldots & TW$^{b}$ \\
LHA\,120-S\,73 &  yes  & yes & \protect{\cite{2012MNRAS.423..284A, 2016A&A...593A.112K}}    &  yes     & 9$\pm$1    & \protect{\cite{1988ApJ...334..639M, 2013A&A...558A..17O, 2016A&A...593A.112K}} & yes & \ldots & \protect{\cite{2017AJ....154..186K}} \\
LHA\,120-S\,89 &  no   & no  & TW$^{b}$       &  no      & \ldots     & \protect{\cite{1988ApJ...334..639M, 2013A&A...558A..17O}} & no & \ldots & TW$^{b}$ \\
LHA\,120-S\,93 &  yes  & no  & TW$^{b}$       &  no      & \ldots     & \protect{\cite{2013A&A...558A..17O}} & no & \ldots & TW$^{b}$ \\
LHA\,120-S\,111  & yes  & yes & \protect{\cite{2012MNRAS.423..284A}}    &  no      & \ldots     & \protect{\cite{1988ApJ...334..639M}} & yes & \ldots & \protect{\cite{1989A&A...220..206Z}} \\
LHA\,120-S\,127  & yes  & yes & \protect{\cite{2007A&A...463..627K, 2012MNRAS.423..284A}}    &  no      & \ldots     & \protect{\cite{1988ApJ...334..639M, 2013A&A...558A..17O}} & no & \ldots & TW$^{b}$ \\
LHA\,120-S\,134  & yes  & yes & \protect{\cite{2012MNRAS.423..284A}}    &  yes     & 15$\pm$2   & \protect{\cite{1988ApJ...334..639M, 2013A&A...558A..17O}} & yes & \ldots & \protect{\cite{1989A&A...220..206Z}} \\
LHA\,120-S\,137  & no   & yes & TW$^{b}$       &  no      & \ldots     & \protect{\cite{2013A&A...558A..17O}} & no & \ldots & TW$^{b}$ \\
\midrule
\multicolumn{10}{c}{Small Magellanic Cloud B[e]SGs}\\
\midrule
LHA\,115-S\,6    & ?    & yes & TW$^{b}$       &  yes     & 12$\pm$2   & \protect{\cite{1989A&A...223..237M, 2013A&A...558A..17O}} & no & \ldots & TW$^{b}$ \\
LHA\,115-S\,18   & yes  & yes & \protect{\cite{2012MNRAS.423..284A}}    &  yes & 20$\pm$5 & \protect{\cite{1996ApJ...470..597M, 2013A&A...558A..17O}} & yes & \ldots & \protect{\cite{1989A&A...220..206Z, 2012MNRAS.427L..80T}}\\
LHA\,115-S\,23$^{c}$   & ?    & no  & TW$^{b}$       &  no      & \ldots     & TW$^{d}$ & no & \ldots & TW$^{b}$ \\
LHA\,115-S\,65   & yes  & yes & \protect{\cite{2010A&A...517A..30K, 2012MNRAS.423..284A}}    &  yes$^{e}$     & 20$\pm$5   & \protect{\cite{2012MNRAS.426L..56O, 2013A&A...558A..17O}} & no & \ldots & TW$^{b}$ \\
\midrule
\multicolumn{10}{c}{Galactic B[e]SGs/B[e]SG Candidates}\\
\midrule
MWC\,137             & no  & no  & \protect{\cite{2017AJ....154..186K, 2018MNRAS.480..320M}}  & yes & 25$\pm$2   & \protect{\cite{2012BAAA...55..123M, 2015AJ....149...13M, 2013A&A...558A..17O}} & no & \ldots & TW$^{b}$  \\
MWC\,349          & yes & \ldots & \protect{\cite{2016MNRAS.456.1424A}}      &  yes   & 4$\pm$1   & \protect{\cite{2000A&A...362..158K, Kraus2019_prep}} & \ldots  & no & \protect{\cite{Kraus2019_prep}}\\
GG\,Car        & yes & no  & \protect{\cite{2018MNRAS.480..320M}}   &  yes & 15$\pm$5   & \protect{\cite{1988ApJ...324.1071M, 1996ApJ...470..597M, 2012BAAA...55..123M, 2013A&A...549A..28K, 2013A&A...558A..17O, 2018MNRAS.480..320M}} & no & \ldots & TW$^{b}$  \\
Hen 3-298            & yes & yes & \protect{\cite{2005A&A...436..653M, 2018MNRAS.480..320M}}   &  yes & 20$\pm$5 & \protect{\cite{2005A&A...436..653M, 2012BAAA...55..123M, 2013A&A...558A..17O, 2018MNRAS.480..320M}} & no & \ldots & TW$^{b}$  \\
CPD-52 9243   & yes & no  & \protect{\cite{2018MNRAS.480..320M}}   &  yes & \ldots    & \protect{\cite{1988ApJ...324.1071M, 2012BAAA...55..123M, 2018MNRAS.480..320M}} & no & yes & TW$^{b}$,\protect{\cite{2015ApJ...800L..20K}} \\
HD\,327083   & yes & no  & \protect{\cite{2018MNRAS.480..320M}}   &  yes & \ldots    & \protect{\cite{2012BAAA...55..123M, 2018MNRAS.480..320M}} & no & yes & TW$^{b}$,\protect{\cite{2015ApJ...800L..20K}}\\
MWC\,300      & no  & yes & TW$^{b}$    &  no  &  \ldots  & \protect{\cite{2012BAAA...55..123M, 2014MNRAS.443..947L}} & no & \ldots &  TW$^{b}$ \\
AS\,381      & no  & no & \protect{\cite{2002A&A...383..171M}}   &  abs  &  ?  & \protect{\cite{2002A&A...383..171M, 2012BAAA...55..123M, 2014MNRAS.443..947L}} & \ldots & \ldots & \ldots \\
CPD-57 2874    & yes & no  & \protect{\cite{2018MNRAS.480..320M}}   &  yes & \ldots    & \protect{\cite{1988ApJ...324.1071M, 2012BAAA...55..123M, 2018MNRAS.480..320M}} & no & yes &  TW$^{b}$,\protect{\cite{2015ApJ...800L..20K}} \\
Hen\,3-938     & yes  & yes & \protect{\cite{2019MNRAS.488.1090C}}    &  \ldots  &  \ldots  &  \ldots & no & \ldots & TW$^{b}$ \\
MWC\,342      & no  & yes &  \protect{\cite{1999A&AS..136...59A, 2016MNRAS.456.1424A}}    &  \ldots   &  \ldots  & \ldots & \ldots & \ldots  & \ldots \\
Hen\,3-303      & no  & no & TW$^{b}$    &  no  &  \ldots  &  \protect{\cite{2005A&A...436..653M}} & no & \ldots & TW$^{b}$ \\
CD-42\,11721      & no  & yes & \protect{\cite{2007MNRAS.377.1343B}} &  no  &  \ldots  & \protect{\cite{2012BAAA...55..123M}} & no & \ldots & TW$^{b}$ \\
HD\,87643      & yes & no  & \protect{\cite{2018MNRAS.480..320M}}  &  yes$^{e}$ & \ldots    & \protect{\cite{2012BAAA...55..123M, 2018MNRAS.480..320M}} & no & \ldots & TW$^{b}$ \\
HD\,62623$^{c}$    & yes & yes & \protect{\cite{2018MNRAS.480..320M}}   &  yes & \ldots    & \protect{\cite{2012BAAA...55..123M, 2018MNRAS.480..320M}} & no & yes & 
TW$^{b}$,\protect{\cite{2015ApJ...800L..20K}}\\
\bottomrule
\multicolumn{10}{p{.98\textwidth}}{{\bf Note:} TW = This work; abs = in absorption; ? = uncertain detection/no value available; \ldots = no information.} \\
\multicolumn{10}{p{.98\textwidth}}{$^{a}$ Refers to the presence of 
[O\,{\sc i}]$\lambda$5577. All sample stars show emission of [O\,{\sc i}]$\lambda\lambda$6300,6364.}\\
\multicolumn{10}{p{.98\textwidth}}{$^{b}$ Based on (unpublished) high-resolution optical spectra 
taken between 2005 and 2017 with FEROS at the MPG 2.2m telescope.} \\ 
\multicolumn{10}{p{.98\textwidth}}{$^{c}$ A[e]SG due to early-A spectral type assignment (\protect{\cite{2008A&A...487..697K, 2010AstBu..65..150C}}).}\\
\multicolumn{10}{p{.98\textwidth}}{$^{d}$ No indication of CO band features seen in a $K$-band 
spectrum (unpublished) taken on 2013 October 20 with OSIRIS at the Southern Astrophysical Research (SOAR) Telescope.}\\ 
\multicolumn{10}{p{.98\textwidth}}{$^{e}$ No CO emission was detected during the observations taken between 1987 and 1989 (\protect{\cite{1988ApJ...324.1071M, 1989A&A...223..237M}}). }\\
\label{tab:CO-O-Ca}
\end{tabular}
\end{table}

\subsection{Disk dynamics and structure}

\begin{figure}[H]
\centering
\includegraphics[width=10cm]{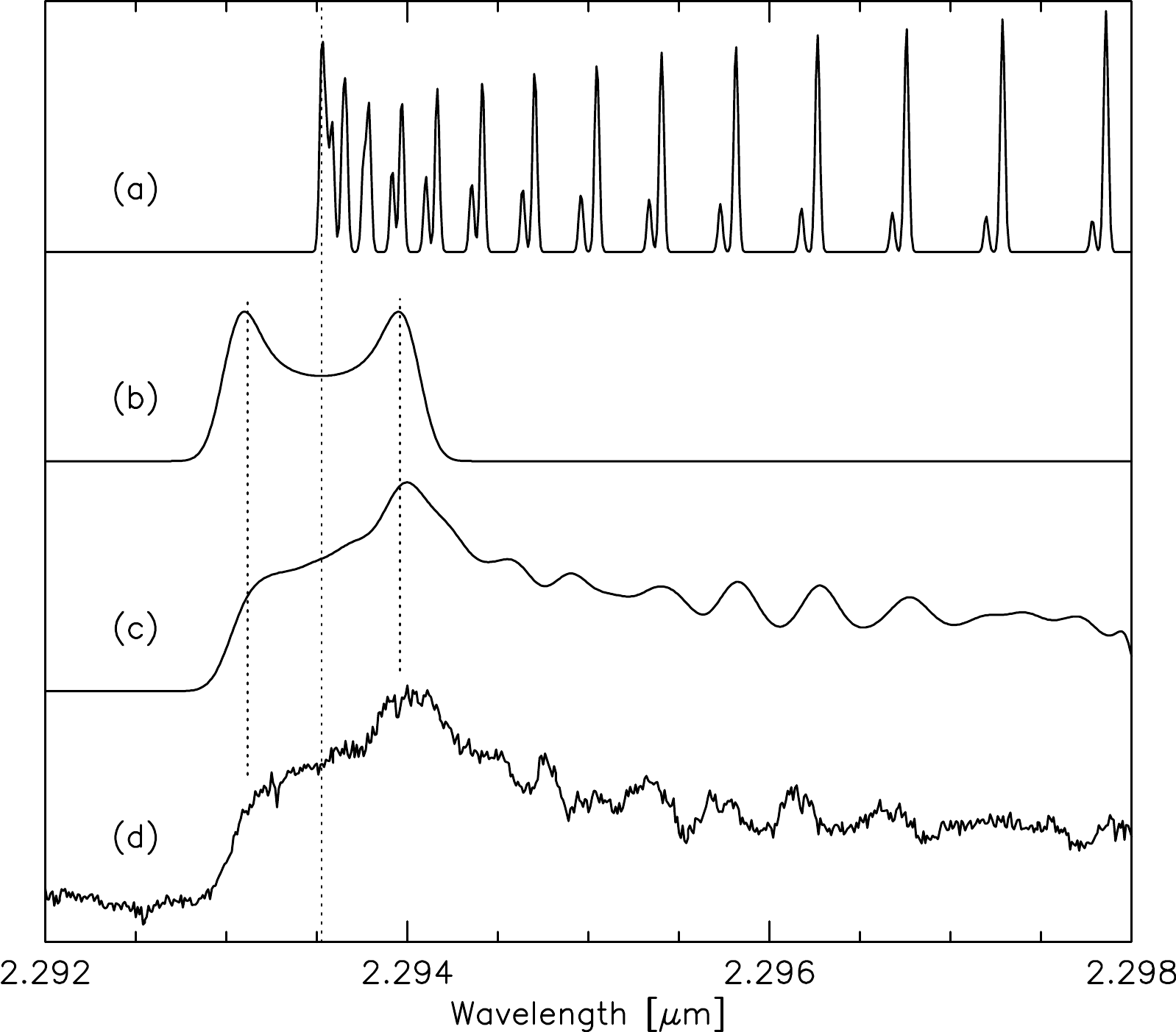}
\caption{Sketch of the generation of the typical CO band head profile. (\textbf{a}) Spectrum around 
the (2-0) band head of the CO first-overtone bands for a hot gas with velocity dispersion of a few 
km\,s$^{-1}$. (\textbf{b}) Profile of a single line from a rotating gas ring with a velocity, 
projected to the line of sight, of 66\,km\,s$^{-1}$ as seen with a spectral resolution of 
6\,km\,s$^{-1}$. (\textbf{c}) Total synthetic CO band head spectrum resulting from the convolution of the band transitions in (\textbf{a}) with the profile of the ring in (\textbf{b}). 
(\textbf{d}) CO band head observations of the Galactic B[e]SG CPD-57\,2874 
(\protect{\cite{2018MNRAS.480..320M}}).}
\label{fig:CO}
\end{figure}

Determination of the kinematics within dense circumstellar environments requires the use of 
reliable tracers. High-resolution near-infrared spectroscopic observations have revealed that the 
band heads of the CO emission from B[e]SGs typically display a characteristic shape, consisting of a 
blue-shifted shoulder and a red-shifted maximum. For the generation of such a band head profile, the 
individual CO rotation-vibration lines, superimposing within the region of the band head, must 
display double-peaked profiles (see Figure\,\ref{fig:CO}). Such line profiles can originate either 
from a circumstellar ring of gas expanding with constant velocity (constant outflow), or from 
rotational motion of a ring of gas around the central object. To discriminate between the two 
scenarios, complementary tracers are needed. 

The SiO band emission seen in four Galactic B[e]SGs displays a similar shape of the band heads. 
Detailed modeling revealed that in each object the SiO bands required a slightly lower value of the 
velocity (\cite{2015ApJ...800L..20K}) than the CO bands. The SiO molecule is less stable than CO, 
meaning that it can form and persist only at lower temperatures. This fact naturally places the region
where SiO molecules are expected to form, and hence the SiO band emitting region, at  
slower orbital velocities and greater distances from the central object.

As CO is the most stable molecule, its formation and emission region marks the inner edge of the
molecular disk. Closer to the star, tracers for the kinematics need to be found from line emission 
of the atomic gas. Here, the lines from forbidden transitions are most suitable, because their
emission is optically thin, so that their profile shapes contain the full velocity information of
their formation region (\cite{2007A&A...463..627K, 2010A&A...517A..30K}). Of particular interest are 
hereby the [O\,{\sc i}] lines, because they are one of the defining characteristics of the B[e] 
phenomenon and hence observed in all B[e]SGs. The ionization potential of O\,{\sc i} is about the same 
as the one for H\,{\sc i}, which means that within the  [O\,{\sc i}] line forming regions, hydrogen 
should be basically neutral as well, restricting the formation region of the [O\,{\sc i}] line emission 
to the neutral regions within the circumstellar disk. While recombination in the equatorial region close 
to the star might be achieved, e.g., with the model of a latitude dependent wind 
(\cite{2003A&A...405..165K, 2006A&A...456..151K, 2008A&A...478..543Z}), the requirement of a hydrogen 
neutral environment severely limits the number of free electrons that will be available to collisionally 
excite the levels within O\,{\sc i} from which the forbidden transitions emerge. Consequently, the [O\,
{\sc i}] lines arise in regions with high total density, but low electron density.

The profiles of the [O\,{\sc i}] lines often display double-peaks, in line with their formation in 
the disk. Typically, the [O\,{\sc i}] $\lambda$5577 line, which arises from a higher level than the  
$\lambda\lambda$6300,6364 lines, is broader, indicating spatially distinct formation regions of the 
emissions, with the [O\,{\sc i}] $\lambda$5577 line being formed at higher velocities and higher 
densities and hence closer to the star than the other two lines.

With the identification of the lines of [Ca\,{\sc ii}] $\lambda\lambda$7291,7324 in the spectra of
numerous B[e]SGs, a further highly valuable tracer for the disk kinematics has been found. These 
lines typically display double-peaked profiles as well, with velocities comparable to or even 
greater than the one traced by the [O\,{\sc i}] $\lambda$5577 line (\protect{\cite{2010A&A...517A..30K, 
2012MNRAS.423..284A, 2018MNRAS.480..320M}}). This implies that they form in the same region, or at least 
very close to each other, which is in agreement with their comparably high critical density. Since the 
[O\,{\sc i}] $\lambda$5577 line is not always detectable, the [Ca\,{\sc ii}] lines thus provide a 
suitable, complementary benchmark for the dynamics within the disks of B[e]SGs.

In summary, the optical and near-infrared spectra provide emission features from several species, 
which are suitable to pin down the kinematics within the disks of B[e]SGs at various distances from 
the star, and Table\,\ref{tab:CO-O-Ca} includes the information on the detection of the individual 
tracers in the MC sample. Based on the physical constraints outlined above, the logical order of the 
appearance of the divers tracers from inside out would be [Ca\,{\sc ii}], [O\,{\sc i}], CO bands and 
SiO bands. The velocity information carried by these species thereby implies a decrease with 
increasing distance from the star. While an equatorial, outwards decelerating outflow might be able 
to explain some of the observed line profiles (\cite{2007A&A...463..627K}), the velocity patterns seem 
to be in better agreement with (quasi-)Keplerian rotation. In this respect it is interesting to note 
that Keplerian rotation has been made directly discernible by means of spectro-interferometric 
observations. The rotational motion of the CO gas has been derived based on the differential phase 
spectrum (\cite{2012A&A...543A..77W, 2012A&A...538A...6W}), and the rotational motion of the ionized gas 
based on the spatially resolved Br$\gamma$ emission (\cite{2011A&A...526A.107M}).   

While, in general, the rotational motion of the material within the circumstellar disks of B[e]SGs 
seems now to be well established, maybe in connection with a (very) slow outflow component 
(\cite{2010A&A...517A..30K}), recent investigations of the spatial distribution of the circumstellar gas 
revealed that it is more likely accumulated in multiple rings, partial rings, and possible spiral arm-
like structures rather than in a smooth disk (\protect{\cite{2016A&A...593A.112K, 2018MNRAS.480..320M, 
2018A&A...612A.113T}}). These rings might result from multiple mass ejection phases caused by 
(pulsational) instabilities acting in the outer layers of these luminous objects (e.g., 
\cite{1993MNRAS.263..375G, 1993MNRAS.264...50K, 1996MNRAS.282.1470G}), or from binary interaction in 
close systems as seems to be the case for some of the Galactic objects 
(\protect{\cite{2018MNRAS.480..320M}}) in which the rings are circumbinary rather than circumstellar. 
Other disk-forming mechanisms that have been proposed in the literature over the years include 
equatorial mass-loss from a critically rotating star (\cite{2014A&A...569A..23K, 2018A&A...613A..75K}), 
the rotationally induced bi-stability mechanism (\cite{2000A&A...359..695P}), the slow-wind solution 
(\cite{2004ApJ...614..929C}), and the combination of the latter two (\cite{2005A&A...437..929C}). For 
an overview including a detailed description of the various models and their limitations, see 
\cite{2017ASPC..508..219K}. 

The circumstellar material of many MC objects appears durable. This is evidenced by their emission 
features and their infrared photometry that both display no considerable variability over several 
decades, in combination with chemically processed dust displaying emission from crystalline silicates 
(\cite{2010AJ....139.1993K}). It is tempting to imagine that in such an environment even minor bodies 
might have formed from the long-lived disk material, creating gaps within the disk in radial direction 
and hence leading to the formation of the presumed ring structures 
(\protect{\cite{2016A&A...593A.112K}}). These minor bodies or possible planets can also stabilize the 
neighboring rings, in analogy to the shepherd moons in planetary systems. But so far, there is 
insufficient observational evidence that might support the validity of such a scenario.

\subsection{Current evolutionary state of B[e]SGs}\label{sect:current_evol}

The formation mechanism of the observed gaseous and dusty rings or disk-like structures around 
B[e]SGs is certainly one of the most important yet unsolved issues. But equally important questions 
arise: What is the evolutionary phase of B[e]SGs? What is their evolutionary connection to other 
evolved massive stars? And exists such a connection at all? While the question on the relation between 
B[e]SGs and other evolved objects is beyond the scope of this review, we will
briefly elucidate on the current knowledge about the evolutionary status of B[e]SGs. Considering the 
MC sample, it is obvious from Fig.\,\ref{fig:HRD-classic} that all objects have evolved off the main 
sequence. Whether this occurred only recently, or whether B[e]SGs might be on a blue loop or blueward 
evolution after having passed through the turning point on the cool edge of their track is still an 
open issue, in particular, since we lack clear methods for age determinations of these emission-line 
objects, which only very rarely are detected in clusters\footnote{Currently only four B[e]SGs are 
reported to be cluster members. These are the two LMC objects LHA\,120-S\,111 in the compact cluster 
NGC\,1994 (\cite{1984Msngr..38...28A}) and LHA\,120-S\,35 in SL482 (\cite{2018A&A...612A.113T}), and the 
two Galactic sources MWC\,137 in SH\,2-266 (\cite{2016A&A...585A..81M}) and Wd1-9 in Westerlund\,1 
(\cite{2005A&A...434..949C, 2013A&A...560A..11C}).}.

\begin{figure}[H]
\centering
\includegraphics[width=13.5cm]{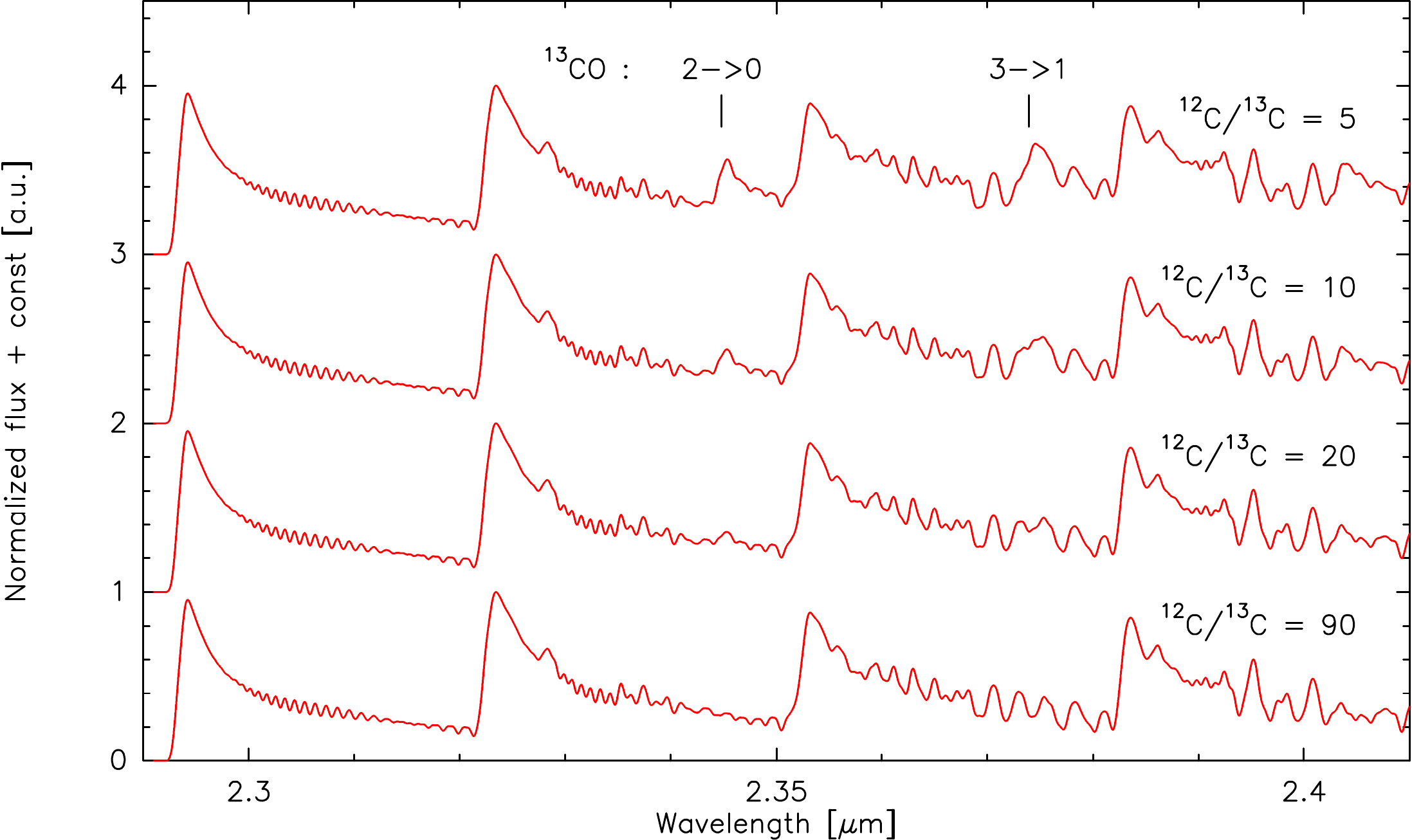} 
\caption{Synthetic spectra of the combined emission from $^{12}$CO and $^{13}$CO for different values
of the $^{12}$C/$^{13}$C ratio. The computations have been performed for the following physical 
parameters: a $^{12}$CO column density of $2\times 10^{21}$\,cm$^{-2}$, a gas temperature of 3000\,K,
a line-of-sight rotational velocity of 66\,km\,s$^{-1}$, and a spectral resolution of 50\,km\,s$^{-1}$.} 
\label{fig:13CO}
\end{figure}

In this regard, the detection of clear signs of the $^{13}$C isotope in form of $^{13}$CO band 
emission from a number of MC B[e]SGs (\protect{\cite{2010MNRAS.408L...6L, 2013A&A...558A..17O}}, see 
Table\,\ref{tab:CO-O-Ca}) was one major step forward. The appearance of these bands has been predicted 
based on theoretical model computations for a variation of the carbon isotope ratio $^{12}$C/$^{13}$C 
(\cite{2009A&A...494..253K}). As surface abundance 
calculations have shown, this ratio will drop during the evolution of massive stars from an initial, 
interstellar value of $\sim$ 90 down to values $<$ 5, depending on the initial mass of the star and 
its initial rotation speed. The surface material enriched in $^{13}$C is transported via winds to the 
environments, where it will cool and condense into $^{13}$CO molecules, whose emission can be observed 
in the $K$-band, together with the emission from the main isotope, $^{12}$CO (see 
Figure\,\ref{fig:13CO}). Hence, the detected amount of $^{13}$CO is a measure for the stellar surface
enrichment in $^{13}$C at the time the material, that is currently traced in the molecular emission, 
has been released from the stellar surface.

From the MC sample of 15 objects, 7 have been found to display CO band emission, and all of them 
display clear indication of enrichment in $^{13}$C (see Table\,\ref{tab:CO-O-Ca}). Interestingly, all
these objects with CO emission cluster in the same region of the HR diagram as indicated by the dotted 
black square in Figure\,\ref{fig:HRD-classic}, i.e., in the luminosity range $\log L/L_{\odot} = 
5.0-5.8$. None of the three most luminous stars (S\,22, S\,111, and S\,127) or of the 4 
low-luminosity objects (S\,23, S\,59, S\,93, and S\,137) displays clear signs of 
CO emission. One outlier in the luminosity domain occupied by the CO emitting B[e]SGs is the star 
S\,89, which also has no detectable CO band emission (\protect{\cite{2013A&A...558A..17O}}).

The absence of measurable CO band emission might have different reasons. Either the intensity of the 
emission is too low to be detectable against the strong near-IR continuum\footnote{This spectral 
region suffers from strong telluric contamination, which is not always easy to remove, so that 
especially weak CO emission features might be hidden within telluric remnants.}, or the density of 
the molecular gas might be very high, resulting in optically thick emission which has no 
characteristic band head structure anymore. Another possibility would be that the CO emission from 
these stars might have variable CO band emission and they have so far always been observed in phases 
of no emission. In this context it is interesting to refer to the SMC object LHA\,115-S\,65, in which 
CO band emission suddenly occurred, while observations taken about nine months earlier did not detect 
any molecular features (\cite{2012MNRAS.426L..56O}). Alternatively, since we now know that the material 
is most probably concentrated in rings, the conditions within the circumstellar environment in terms of 
density and temperature might not be favorable for the excitation of the first-overtone bands. For 
those stars, observations in the spectral region of the fundamental bands might therefore be a 
possibility to search for cooler CO gas. 

\begin{figure}[H]
\centering
\includegraphics[width=10cm]{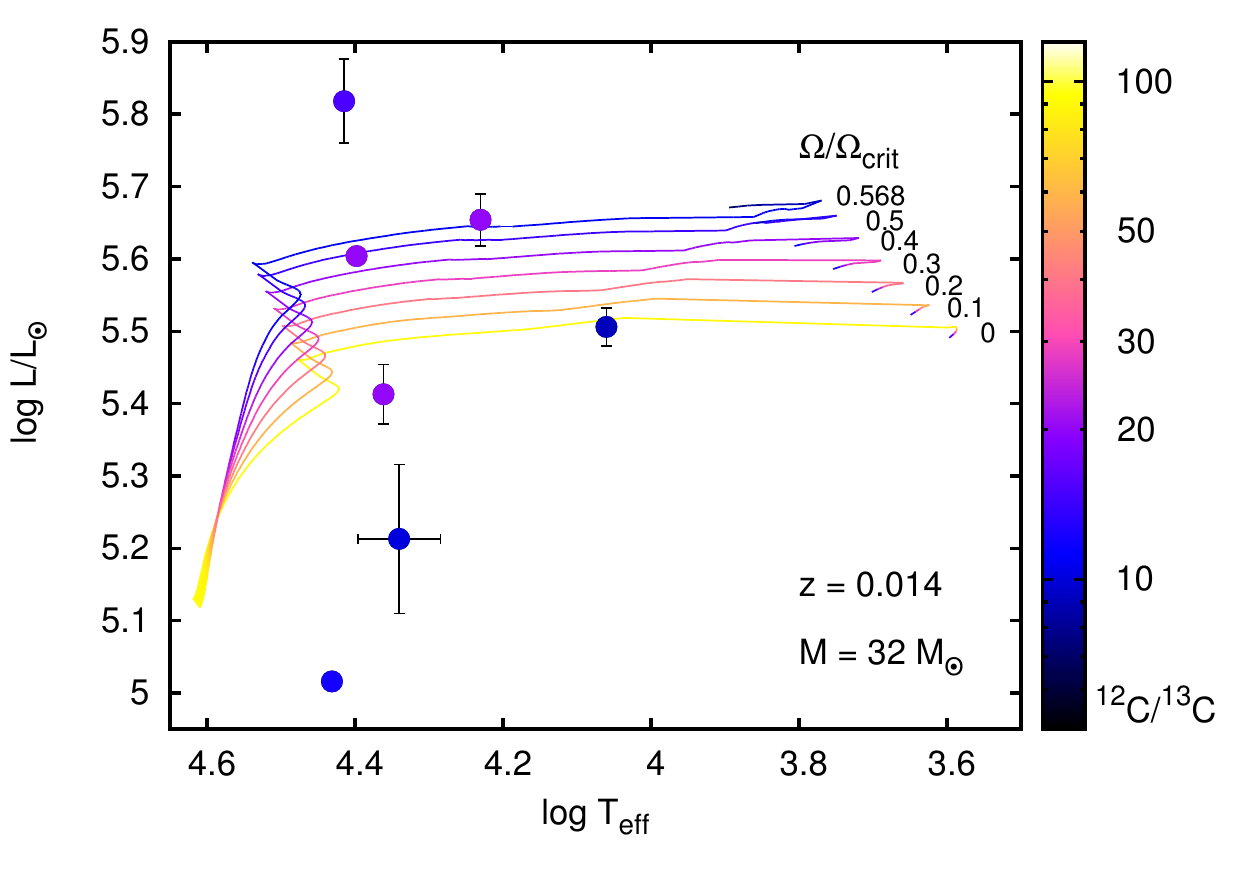}
\caption{Evolution of the $^{12}$C/$^{13}$C isotope ratio along the solar metallicity tracks of a star
with initial mass of 32\,M$_{\odot}$ and initial rotation speeds 
$v/v_{\rm crit}$ ranging from 0 to 0.4 (corresponding to 
$\Omega/\Omega_{\rm crit} = 0.0; 0.1; 0.2; 0.3; 0.4; 0.5; 0.568$). The 
individual tracks have been obtained from the interpolation tool SCYCLIST provided by the Geneva 
group. For clarity of the plot, we truncated the evolutionary tracks within the red supergiant 
regions. Included are the positions of the MC B[e]SG sample from Table\,\ref{tab:CO-O-Ca} with known
$^{12}$C/$^{13}$C ratio, following the same color coding as for the tracks. The Galactic 
objects are excluded due to their highly uncertain luminosities. Depending on the 
initial rotation speed of the star, the observed ratio can be reached either in the pre-RSG (moderate 
rotator) or post-RSG (slow rotator) phase.}
\label{fig:surface_enrich}
\end{figure}

The measured $^{12}$C/$^{13}$C isotope ratios of the MC B[e]SGs are all very similar, spreading from  
9 to 20. This might point towards a similar formation history of the circumstellar material, i.e. a 
similar phase in the evolution (considering they are single stars or at least unaffected by a possible 
companion) when the enriched material was ejected. Considering that stars are typically born with some 
intrinsic rotation velocity, rotational mixing in combination with enhanced mass-loss may drive the 
enrichment of the stellar surface with $^{13}$C already in early stages of the evolution of massive 
stars. This can be seen in Figure\,\ref{fig:surface_enrich}, where the evolutionary tracks of a 
32\,M$_{\odot}$ star with solar metallicity (\cite{2012A&A...537A.146E}) and for a variety of initial 
rotation velocities are shown. The covered rotation speeds spread from $v/v_{\rm crit} = 0$ to 
$v/v_{\rm crit} = 0.4$, which correspond to values of $\Omega/\Omega_{\rm crit}$ from 0 to 0.568. The 
interpolation of the tracks has been performed with the SCYCLIST 
tool\footnote{https://www.unige.ch/sciences/astro/evolution/en/database/syclist/} provided by the 
Geneva group. 

The color coding along the tracks refers to the values of the $^{12}$C/$^{13}$C isotope ratio on the 
stellar surface. Also included in Figure\,\ref{fig:surface_enrich} are the positions of the seven MC 
B[e]SGs with known values of the $^{12}$CO/$^{13}$CO isotope ratios. Their colors correspond to the 
same color coding as the evolutionary tracks. Obviously, the observed ratios for the sample stars
might be reached either along or after the main-sequence evolution for stars rotating initially with
rates $\Omega/\Omega_{\rm crit} \ge 0.3$. But they might also be reached during or after the red 
supergiant stage for stars rotating initially with rates smaller than $\Omega/\Omega_{\rm crit} \le 
0.3$. As we do not know the initial rotation speeds of the progenitor stars of these B[e]SGs, the 
measured values of the $^{12}$CO/$^{13}$CO isotope ratio alone cannot solve the issue with the 
current evolutionary state of the objects. 

If B[e]SGs represent a specific phase in the evolution of massive stars, then these objects should
also exist in other environments with high content of massive stars. Searching for representatives  
of B[e]SGs in other galaxies and studying their properties and number statistics at various 
metallicities might help to unveil their disk/ring formation mechanism, to pin down 
their evolutionary phase (pre- versus post-RSG), and to set constraints on the 
evolution of massive stars in general.


\section{Identification of B[e] supergiants in the Local Universe}\label{sect:method}

Identifying and classifying B[e]SGs is a tedious job, whether in the Milky Way, where
we face large amounts of foreground extinction and the issue with the often unknown distances, or in
other galaxies, where we have strong contamination with foreground sources, often crowded regions, and 
the faintness of the objects. While the first B[e]SGs have been found rather accidentally, nowadays 
dedicated surveys can make use of the established classification criteria. As mentioned in 
Section\,\ref{sect:B[e]SGs}, there are basically four characteristics a star should fulfill to be 
classified as B[e]SGs (\cite{Lamers1998, 2006ASPC..355..135Z}). It should display:
\begin{itemize}[leftmargin=*,labelsep=5.8mm]
\item	A spectral-type B (with extensions to late-O and early-A types), as evidenced from the hot 
underlying continuum or from photospheric absorption features\footnote{We would like to stress that 
the high-luminosity B[e]SGs barely display photospheric absorption lines whereas the low-luminosity 
B[e]SGs typically do.}, and a luminosity higher than $\log L/L_{\odot} \sim 4$. 
\item	Very intense Balmer lines dominating the optical spectrum with an equivalent width of H$\alpha$ reaching up to 1000\,\AA.
\item	A spectrum indicating the hybrid character of the environment, consisting of a hot, fast 
line-driven wind coexisting with a cool, slow component from which the forbidden emission lines from 
neutral and low-ionized metals originate ([O\,{\sc i}], [Fe\,{\sc ii}]). 
\item   Strong near- and mid-infrared excess emission, indicative of hot circumstellar dust. 
\end{itemize}
Despite these defining criteria, the identification of extragalactic B[e]SGs is not as 
straightforward, because suitable candidates need first to be found based on other means. These 
candidates can then be further investigated to search for these, mostly spectroscopic, 
characteristics.

A suitable approach is to search for B[e]SG candidates among the luminous, blue objects identified in 
imaging surveys that have been performed for several galaxies over the past $\sim$30 years. For 
instance, the early photographic and photometric surveys of M31 (\cite{1988A&AS...76...65B, 
1986AJ.....92.1303M}) and M33 (\cite{1984PhDT........29F}) revealed (amongst many other objects) the 
most luminous hot stars and the brightest blue supergiants, which could then be studied 
spectroscopically to obtain indications for their possible nature (e.g. \cite{1990AJ.....99...84H}). A 
milestone for the identification of evolved massive stars was certainly the Local Group Galaxies Survey 
(LGGS) project (\cite{2006AJ....131.2478M, 2007AJ....133.2393M}), which resulted in the discovery of 
numerous putative emission-line stars in M31, M33, and seven more dwarf galaxies. Spectroscopic 
follow-ups were used to sort out H\,{\sc ii} regions, and to match the remaining objects with the 
various known categories of evolved massive stars. A major result of this survey was the 
identification of numerous objects that were dubbed  as luminous blue variable (LBV) 
candidates (\cite{2007AJ....134.2474M, 2016AJ....152...62M}), based on the appearance of their blue 
emission-line or P\,Cygni-type spectra that resemble  confirmed LBVs in quiescence. 
Since the blue optical spectra of LBVs in their quiescence state display a number of common 
characteristics with B[e]SGs (\cite{2000AJ....119.2214M, 2007AJ....134.2474M, 2012A&A...541A.146C}), it 
was expected that this sample contains a number of B[e]SGs. 

Surprisingly, when analyzed in more detail (\cite{2012A&A...541A.146C, 2014ApJ...790...48H, 
2017ApJ...836...64H}), this bunch of newly identified LBV candidates in M31 and M33 turned out to be a 
mixed bag containing not only B[e]SGs candidates, but also so-called Fe\,{\sc ii} emission stars (with 
neither [O\,{\sc i}] nor [Fe\,{\sc ii}] emission and lacking warm dust), and warm hypergiants (with lots 
of dust, possible [O\,{\sc i}] and/or [Fe\,{\sc ii}] but of spectral type A-F), with only a few objects 
left to be considered as LBV candidates (with no [O\,{\sc i}] emission and lacking hot 
dust). 

\begin{figure}[H]
\centering
\includegraphics[width=15cm]{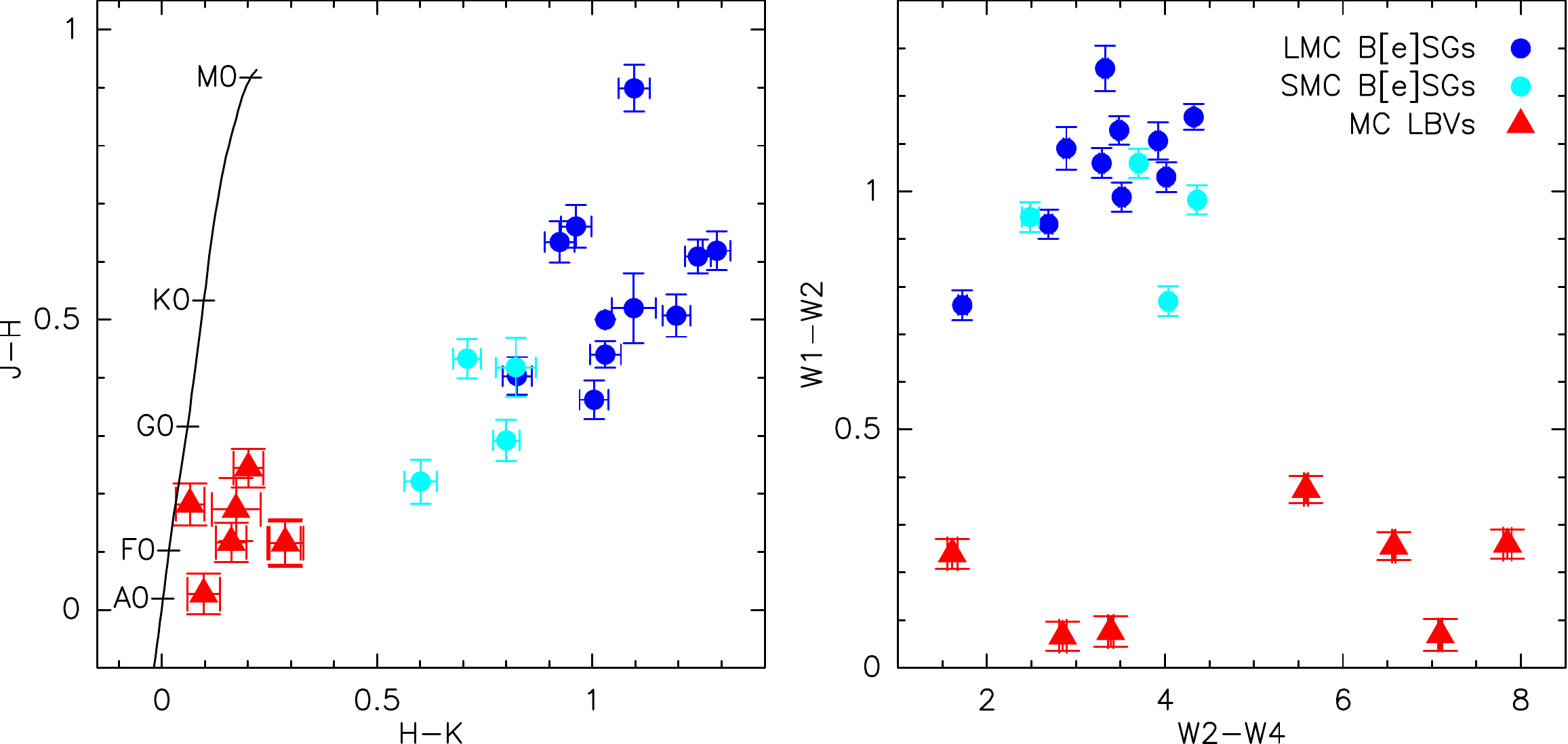}
\caption{Demonstration of the separation of the B[e]SGs from the quiescent LBVs 
within the near-IR ($J-H$ versus $H-K$ diagram, left panel) and the WISE diagram ($W1-W2$ versus $W2-W4$, 
right panel). Shown are the positions of the classical MC B[e]SG sample and of the MC LBV sample. IR 
colors of the objects are provided in Table\,\ref{tab:JHK}. The solid line represents the positions of 
regular supergiants with empirical colors taken from \cite{2011ApJS..193....1W} for solar metallicity 
stars.} 
\label{fig:JH-HK}
\end{figure}

As both, LBVs and B[e]SGs, are luminous blue supergiants, they share the same optical colors. Hence, one step 
to distinguish these two groups of objects is to inspect their location in infrared color-color 
diagrams (e.g. \cite{1986A&A...163..119Z, 1995A&A...302..409G, 2012A&A...541A.146C, 2013A&A...558A..17O, 
2017ApJ...836...64H}). The hot ($\sim$ 1000\,K) circumstellar dust of B[e]SGs results in significantly 
increased near-IR emission. On the other hand, LBVs can be associated to cold, dusty 
environments such as the circumstellar shells recently discovered around many LBVs 
and LBV candidates with the Spitzer Space Telescope at 24\,$\mu$m (e.g. 
\cite{2010MNRAS.405.1047G, 2010AJ....139.2330W}). The separation of B[e]SGs from 
quiescent LBVs, based on their diverse IR properties, is demonstrated in Figure\,\ref{fig:JH-HK} for 
the known samples of MC objects, limiting to the confirmed and generally accepted LBVs in the LMC 
(\cite{2018AJ....156..294A}) and including one confirmed object from the SMC (R40, 
\cite{2018A&A...613A..33C}). Shown are two different color-color diagrams ($J-H$ versus $H-K$ and 
$W1-W2$ versus $W2-W4$). The IR colors of the objects are listed in Table\,\ref{tab:JHK}. They result 
from the $JHK$-band magnitudes obtained from the 2MASS point source 
catalog\footnote{http://cdsarc.u-strasbg.fr/viz-bin/cat/II/246} (\cite{2003yCat.2246....0C}), and from 
the mid-IR magnitudes ($W1, W2, W4$) collected with the Wide-field Infrared Survey 
Explorer\footnote{http://cdsarc.u-strasbg.fr/viz-bin/cat/II/311} (WISE, \cite{2012yCat.2311....0C}). 
From the many possibilities of near- and mid-IR color-color diagrams, these two show the clearest 
separation between the two groups of objects.

Another clear distinctive feature between B[e]SGs and LBVs is the S\,Dor cycle of the latter, while 
B[e]SGs are typically not undergoing this type of variability. However, to identify such S\,Dor 
excursions of the stars within the HR diagram is a time-consuming (though important) task, because it 
requires regular monitoring of the whole sample.

In addition, dedicated spectroscopic observations are required to search for the characteristic 
forbidden emission lines of [O\,{\sc i}] and possible [Ca\,{\sc ii}] in the red portion of the optical 
spectra, and to search for possible molecular emission of CO in their near-IR spectra, because LBVs 
typically do not show these sets of forbidden lines and CO band emission\footnote{One exception to 
this rule is the LBV star HR\,Car, which occasionally showed CO band emission related to phases when 
the star was dimmer (\cite{1997ASPC..120...20M}).} (\cite{1988ApJ...334..639M, 1988ApJ...324.1071M, 
2013A&A...558A..17O}).

\begin{table}[H]
\caption{IR colors of Magellanic Cloud B[e]SGs and LBVs.}
\centering
\begin{tabular}{lcccccl}
\toprule
\textbf{Object}	& \sl{\textbf{J-H}} & \sl{\textbf{H-K}}  & \sl{\textbf{W1-W2}} & \sl{\textbf{W2-W4}} & \textbf{Class}\\
\midrule
\multicolumn{6}{c}{Large Magellanic Cloud}\\
\midrule
LHA\,120-S\,12   & 0.634$\pm$0.035 & 0.924$\pm$0.035 & 0.761$\pm$0.031 &  1.723$\pm$0.051 & B[e]SG\\
LHA\,120-S\,22   & 0.619$\pm$0.033 & 1.289$\pm$0.033 & 1.258$\pm$0.047 &  3.328$\pm$0.032 & B[e]SG\\
LHA\,120-S\,35   & 0.520$\pm$0.060 & 1.096$\pm$0.051 & 1.156$\pm$0.027 &  4.323$\pm$0.021 & B[e]SG\\
LHA\,120-S\,59   & 0.661$\pm$0.037 & 0.962$\pm$0.035 & 0.988$\pm$0.031 &  3.514$\pm$0.033 & B[e]SG\\
LHA\,120-S\,73   & 0.403$\pm$0.033 & 0.825$\pm$0.033 & 1.030$\pm$0.031 &  4.015$\pm$0.027 & B[e]SG\\
LHA\,120-S\,89   & 0.440$\pm$0.022 & 1.030$\pm$0.036 & 1.059$\pm$0.031 &  3.289$\pm$0.026 & B[e]SG\\
LHA\,120-S\,93   & 0.507$\pm$0.036 & 1.195$\pm$0.033 & 1.128$\pm$0.030 &  3.486$\pm$0.025 & B[e]SG\\
LHA\,120-S\,111$^{a}$  & 0.500$\pm$0.000 & 1.030$\pm$0.000 &    \ldots &           \ldots & B[e]SG\\
LHA\,120-S\,127  & 0.362$\pm$0.033 & 1.004$\pm$0.033 & 1.106$\pm$0.039 &  3.922$\pm$0.028 & B[e]SG\\
LHA\,120-S\,134$^{b,c}$   & 0.609$\pm$0.030 & 1.245$\pm$0.030 & 1.090$\pm$0.045 &  2.892$\pm$0.035 & B[e]SG\\
LHA\,120-S\,137  & 0.899$\pm$0.040 & 1.097$\pm$0.037 & 0.931$\pm$0.030 &  2.691$\pm$0.045 & B[e]SG\\
LHA 120-S 96 (= S\,Dor)& 0.173$\pm$0.054 & 0.172$\pm$0.057 & 0.374$\pm$0.028 & 5.586$\pm$0.022 & LBV\\
LHA 120-S 116 (= R110) & 0.116$\pm$0.034 & 0.161$\pm$0.036 & 0.076$\pm$0.032 & 3.391$\pm$0.034 & LBV\\
LHA 120-S 128 (= R127) & 0.115$\pm$0.037 & 0.286$\pm$0.036 & 0.255$\pm$0.029 & 6.573$\pm$0.021 & LBV\\
LHA 120-S 155 (= R71)  & 0.027$\pm$0.035 & 0.097$\pm$0.038 & 0.069$\pm$0.033 & 7.095$\pm$0.024 & LBV\\
CPD-69 463 (= R143)    & 0.244$\pm$0.033 & 0.201$\pm$0.035 & 0.259$\pm$0.030 & 7.851$\pm$0.049 & LBV\\
LHA\,120-S\,83 (= Sk $-$69 142a) & 0.115$\pm$0.041 & 0.287$\pm$0.041 & 0.239$\pm$0.031 & 1.609$\pm$0.055 & LBV\\
\midrule
\multicolumn{6}{c}{Small Magellanic Cloud}\\
\midrule
LHA\,115-S\,6$^{b}$    & 0.433$\pm$0.034 & 0.709$\pm$0.0319 & 0.769$\pm$0.031 & 4.037$\pm$0.035 & B[e]SG\\
LHA\,115-S\,18$^{b,c}$   & 0.418$\pm$0.050 & 0.822$\pm$0.0460 & 1.059$\pm$0.030 & 3.705$\pm$0.031 & B[e]SG\\
LHA\,115-S\,23   & 0.221$\pm$0.038 & 0.601$\pm$0.0382 & 0.946$\pm$0.031 & 2.484$\pm$0.096 & B[e]SG\\
LHA\,115-S\,65   & 0.292$\pm$0.035 & 0.800$\pm$0.0311 & 0.982$\pm$0.030 & 4.363$\pm$0.027 & B[e]SG\\
LHA 115-S 52 (= R40) & 0.182$\pm$0.036 & 0.065$\pm$0.033 & 0.066$\pm$0.031 & 2.851$\pm$0.043 & LBV\\
\bottomrule
\multicolumn{6}{p{.9\textwidth}}{{\bf Note:} IR photometry for all objects is taken from the 2MASS 
point source catalog ($J, H, K$, \cite{2003yCat.2246....0C}), except for the stars LHA\,120-S\,111 (\cite{1988ApJ...334..639M}) and LHA\,120-S\,89 (\cite{2009AJ....138.1003B}), and from the WISE All-Sky Data Release ($W1, 
W2, W4$, \cite{2012yCat.2311....0C}). }\\
\multicolumn{6}{p{.9\textwidth}}{{$^{a}$}Despite of the lack of WISE colors, the presence of warm dust 
is proven by its IR excess seen in the {\sl Spitzer} data, and its IR spectrum that looks like a twin of the one of LHA\,120-S\,73 (see \cite{2010AJ....139.1993K}).}\\
\label{tab:JHK}
\end{tabular}
\end{table}


\section{A census of B[e]SGs}\label{sect:census}

With clearly defined classification characteristics and the proper observational tools at hand, the 
massive star population within the local Universe can be scanned for suitable B[e]SG candidates. At 
the moment of writing this review, this is still an ongoing project that requires patience and 
sufficient telescope time at both, optical and infrared facilities. Nevertheless, many new, 
particularly extragalactic B[e]SG star discoveries were reported in the literature within the past 20 
years. The aim of this section is, therefore, to take a closer and critical look at the suggested 
B[e]SG candidates in order to sort out possible misclassified objects, to check what type of 
observations are still missing for unambiguous classification of the candidates, and to compile 
updated lists of confirmed B[e]SGs for the galaxies with a reported B[e]SG population. Starting point 
for this investigation are the Magellanic Clouds (Section\,\ref{sect:MagellanicClouds}), moving 
further out into the Local Group (Section\,\ref{sect:LocalGroup}) and beyond 
(Section\,\ref{sect:BeyondLG}), before we finally return to the Milky Way 
(Section\,\ref{sect:MilkyWay}). 

The samples in each galaxy are presented in tables, which follow the same structure. The objects are 
listed under a (if possible) homogeneous SIMBAD identifier in column 1, reference(s) for the B[e]SG 
classification of the stars follow in column 2. Where available, $E(B-V)$ values and their references
are provide in columns 3 and 4, and the four colors ($J-H$, $H-K$, $W1-W2$, and $W2-W4$) are given 
along with their errors in the last four columns. The tables are furthermore organized such that in the 
top part appear the confirmed B[e]SGs. These are stars that fulfill all classification criteria. In the 
middle part of each table stars with uncertain or controversial classification are listed with their 
names in parentheses. These are objects that lack one or more of the classification criteria due to 
incomplete observational data sets. Also included are here objects for which different research teams 
find controversial results so that clarification is needed. In the bottom of each table erroneously 
classified objects are gathered with their names in italic and within parentheses. These are stars for 
which observational evidence (e.g. a specific color) excludes them from belonging to the class of 
B[e]SGs.

\subsection{Magellanic Clouds}\label{sect:MagellanicClouds}

As mentioned in Section\,\ref{sect:B[e]SGs}, the "classical" sample of B[e]SGs resides within the 
Magellanic Clouds. The first such object, for which the hybrid character was reported, was the LMC star 
RMC 126 (LHA 120-S 127, \cite{1985A&A...143..421Z}), and soon after followed the identification of ten 
more B[e]SGs in the LMC and four in the SMC (\cite{1989ASSL..157..295S, 1986A&A...163..119Z, 
1989A&A...220..206Z, 1992A&A...260..205Z, 1996A&A...309..505Z, 1995A&A...302..409G}, see 
Table\,\ref{tab:logTL}) based on a dedicated search for similar objects. Since then, a few more stars 
have been suggested as B[e]SGs candidates. These are presented and discussed in the following. The 
classical sample of B[e]SGs in the MCs, which fulfill all classification criteria, has already been listed in Table\,\ref{tab:logTL}, with 
their IR colors provided in Table\,\ref{tab:JHK}.

\subsubsection{Large Magellanic Cloud}
 
\begin{figure}[H]
\centering
\includegraphics[width=11.5cm]{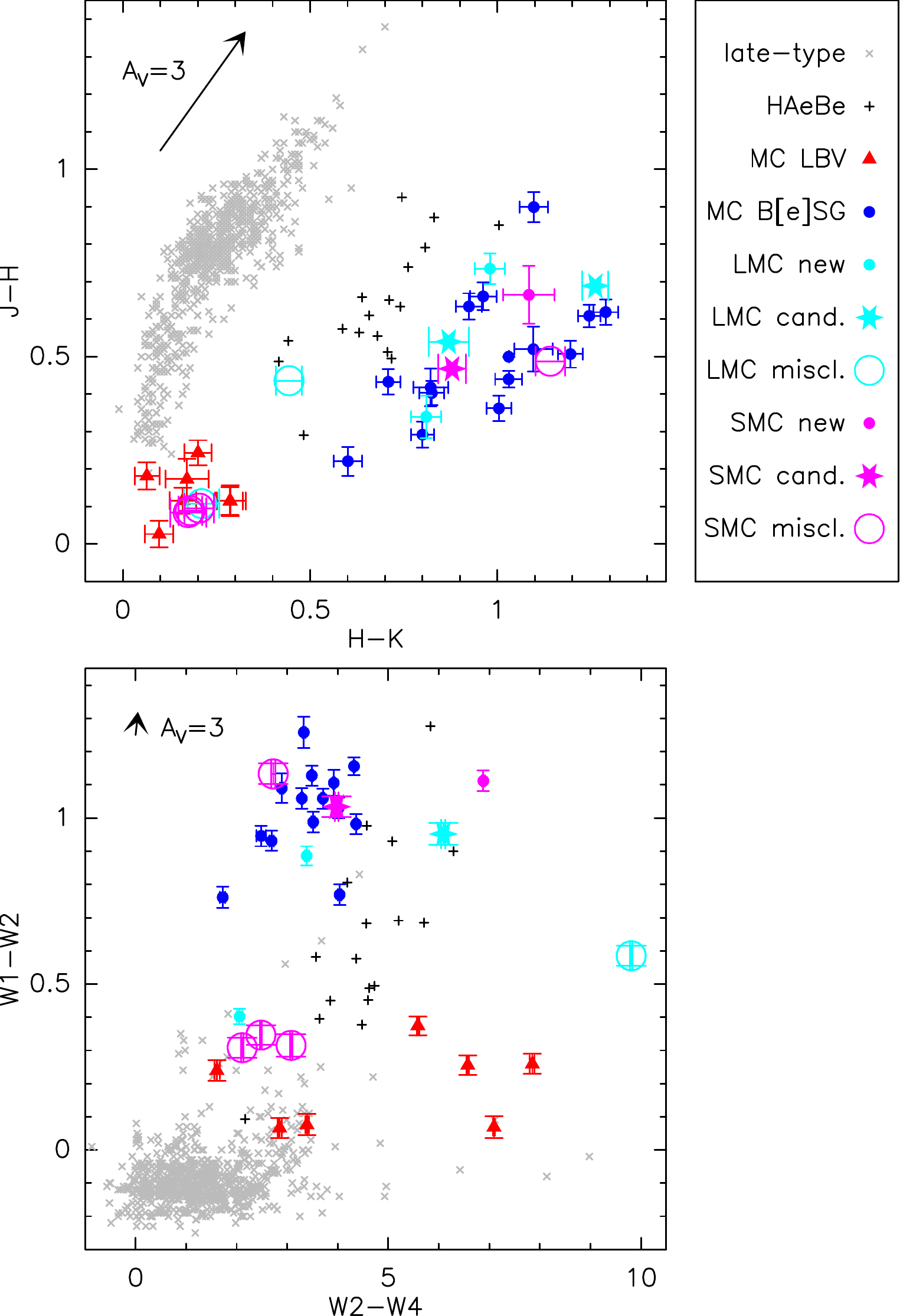}
\caption{Location of the new LMC (light blue) and SMC (purple) samples with respect 
to those MC B[e]SGs that meet all the required classification criteria (dark blue) 
and LBVs (red triangles) in the near-IR (top panel) and the WISE diagram (bottom panel).  
Filled circles are used for confirmed B[e]SGs, filled stars for candidates, and 
empty circles for misclassified objects. Also included are a sample of late-type stars 
and supergiants in the MCs (small gray crosses, from \cite{2015A&A...578A...3G}) and a sample of 
(dereddened, see text) Galactic HAeBe stars (black plus signs, from \cite{2004AJ....127.1682H}).
The arrow in each panel indicates the direction of the reddening, and their length complies 
with a value of $A_{\rm V} = 3$. As not for all MC objects the color excess is known, no extinction 
correction has been applied. But the MC stars have in general relatively small color excess values
(see Tables\,\ref{tab:JHK-MCnew} and \ref{tab:SMC}) which would shift them only marginally in the 
diagrams.}
\label{fig:color-color-MCnew}
\end{figure}

New B[e]SG candidates have been found either by dedicated searches, e.g. from cross-matching catalogs 
of emission-line stars with near-IR catalogs (e.g \cite{2011IAUS..272..260M}), or more serendipitously 
as a by-product of deep spectroscopic surveys of specific regions, such as the VLT-FLAMES Tarantula 
Survey (VFTS, \cite{2011A&A...530A.108E, 2015A&A...574A..13E, 2014A&A...564L...7K}) that was devoted to 
the 30 Doradus starburst region. Also surveys for other purposes, such as the search for post-asymptotic 
giant branch stars (\cite{2015MNRAS.454.1468K}), resulted in new B[e]SG candidates\footnote{I would like 
to point out that from the proposed 12 newly discovered B[e] stars only one was found to be a 
supergiant (see Table\,\ref{tab:JHK-MCnew}). The others have been carefully inspected in collaboration 
with Devika Kamath, and they did not fulfill the requirements. These results are yet unpublished.}.
Seven new B[e]SG objects have been proposed in total, of which only two fulfill the criteria for 
B[e]SGs, two are considered candidates, and three appear to be misclassified. All seven stars are 
listed in Table\,\ref{tab:JHK-MCnew}. Their locations in the two color-color diagrams are shown
in Figure\,\ref{fig:color-color-MCnew} in comparison with the confirmed B[e]SGs and LBVs from 
Table\,\ref{tab:JHK}. 

A sample of confirmed MC late-type stars and supergiants (\cite{2015A&A...578A...3G}) is included 
in these color-color diagrams. These serve as reference for objects that might have a late-type 
companion. The outliers of the late-type stars, especially in the WISE diagram, are objects with high 
mass-loss.

Also shown is a sample of Galactic Herbig AeBe (HAeBe) stars (\cite{2004AJ....127.1682H}). Only stars 
with known extinction values and solid magnitudes in all four bands were selected (rejecting objects 
with reported contamination). As some of these pre-main sequence objects suffer from very high 
extinction ($A_{\rm V} > 5$\,mag), their colors were corrected using an $R_{\rm V}$-dependent 
extinction law (\cite{1989ApJ...345..245C}) with $R_{\rm V} = 5.0$ which has been found to be reasonable 
for HAeBe stars (\cite{2004AJ....127.1682H}). These pre-main sequence stars were included to check for 
possible misclassification of objects that have a proposed luminosity ranging around the lower limit 
for B[e]SGs, because this luminosity range is shared by the most massive HAeBes. In the near-IR 
diagram, the HAeBes populate a stripe that appears to be parallel and seems to connect seamlessly to 
the region occupied by the classical B[e]SGs. In the WISE diagram, the HAeBes also seem to populate a 
stripe adjacent to the B[e]SG domain. 

In the following, the reasons for classification of the new LMC objects as either confirmed or 
candidate, or for rejecting them as B[e]SGs are briefly presented.

\paragraph{\bf LHA\,120-S 165}
This object was first listed as a candidate young stellar object (\cite{2008AJ....136...18W}), but was 
later classified as possible B[e]SG, based on its optical spectrum displaying all characteristic 
emission features (SSTISAGEMC J052747.62-714852.8, \cite{2015MNRAS.454.1468K}). Its optical and infrared 
brightness together with its position in both color-color diagrams support this classification.

\paragraph{\bf ARDB 54}
Recent analysis of ARDB 54 revealed that it belongs to the (so far only few) A[e]SGs, being the first
of its kind in the LMC (\protect{\cite{2019MNRAS.488.1090C}}). Its position in the near-IR diagram 
supports this classification, although the object is displaced from the region of classical B[e]SGs in 
the WISE diagram, where it is located closer to the LBV region. Its luminosity of $\log L/L_{\odot} 
\simeq 4.4$ (\protect{\cite{2019MNRAS.488.1090C}}) is too low to be considered as LBV candidate, and a 
bit too high to be considered as HAeBe star. From the latter ARDB 54 is also offset in both color-color 
diagrams, so that a classification as A[e]SG seems to be the most reasonable, despite its exposed 
location in the WISE diagram, whose cause should be examined more closely.

\begin{table}[H]
\caption{Confirmed and candidate B[e]SGs in the LMC. Misclassified objects are listed in the bottom part of the table.}
\centering
\begin{tabular}{lccccccc}
\toprule
\textbf{Object}	& \textbf{Ref.} & \sl{\textbf{E(B-V)}} & \textbf{Ref.} & \sl{\textbf{J-H}} & \sl{\textbf{H-K}}  & \sl{\textbf{W1-W2}} & \sl{\textbf{W2-W4}}  \\
\midrule
\multicolumn{8}{c}{Confirmed B[e]SGs}\\
\midrule
LHA\,120-S\,165  & \protect{\cite{2015MNRAS.454.1468K}} & \ldots & \ldots & 0.735$\pm$0.041 & 0.98$\pm$0.040 & 0.886$\pm$0.029 & 3.385$\pm$0.037 \\
ARDB 54 & \protect{\cite{2014A&A...568A..28L, 2019MNRAS.488.1090C}} & 0.11 & \protect{\cite{2019MNRAS.488.1090C}}  &  0.340$\pm$0.057 & 0.810$\pm$0.04 & 0.402$\pm$0.023 & 2.057$\pm$0.054 \\
\midrule
\multicolumn{8}{c}{Uncertain or controversial classification}\\
\midrule
(VFTS 1003)         & \protect{\cite{2011A&A...530A.108E}} & \ldots & \ldots &  0.540$\pm$0.094 & 0.87$\pm$0.0539 &  \ldots & \ldots  \\
(VFTS 822)  &  \protect{\cite{2014A&A...564L...7K, 2015A&A...574A..13E}} & 0.56 & \protect{\cite{2014A&A...564L...7K}} &  0.689$\pm$0.039 & 1.261$\pm$0.035 & 0.952$\pm$0.033 &  6.09$\pm$0.046 \\
\midrule
\multicolumn{8}{c}{Erroneous classification}\\
\midrule
({\sl VFTS 698})  & \protect{\cite{2012A&A...542A..50D}} & 0.6 & \protect{\cite{2012A&A...542A..50D}} &  0.436$\pm$0.030 & 0.444$\pm$0.034 & 0.585$\pm$0.030 & 9.81$\pm$0.023 \\
({\sl [L72] LH 85-10}) & \protect{\cite{2000AJ....119.2214M}} & \ldots & \ldots &   0.108$\pm$0.036 & 0.211$\pm$0.048 & \ldots & \ldots  \\
({\sl NOMAD1 } &  &  &  &   & &  &  \\
{\sl  0181-0125572})  & \protect{\cite{2014A&A...568A..28L}} & \ldots & \ldots &  \ldots$^{a}$ & \ldots$^{a}$ & \ldots & \ldots \\
\bottomrule
\multicolumn{8}{p{.9\textwidth}}{{\bf Note:} IR photometry for 
all objects is taken from the 2MASS point source catalog ($J, H, K$, \cite{2003yCat.2246....0C}) and from the WISE All-Sky Data Release ($W1, W2, W4$, \cite{2012yCat.2311....0C}).}\\
\multicolumn{8}{p{.9\textwidth}}{$^{a}$ The JHK magnitudes listed in SIMBAD were mistakenly taken from 
the paper of \protect{\cite{2014A&A...568A..28L}}, but these belong to the star LHA\,120-S\,165.}
\label{tab:JHK-MCnew}
\end{tabular}
\end{table}

\paragraph{\bf (VFTS 1003)}
This star was found from the VLT-FLAMES Tarantula Survey (\protect{\cite{2011A&A...530A.108E}}). In high 
angular-resolution near-IR images it appears as a single, isolated object (\cite{2010MNRAS.405..421C}).
The star's blue emission-line spectrum resembles closely the one of the Galactic B[e]SG GG\,Car.  
VFTS 1003 was suggested to be either a Herbig B[e] star or a B[e] supergiant 
(\protect{\cite{2011A&A...530A.108E}}). 
From its position in the near-IR diagram, a supergiant classification seems more likely. But the lack 
of WISE photometry allows to assign VFTS 1003 only a candidate status. In addition, for a definite 
B[e] classification, red optical spectra are needed, because the optical spectra reported in the 
literature do not cover the region of the [O\,{\sc i}] lines.


\paragraph{\bf (VFTS 822)} 

This late B-type star (\protect{\cite{2014A&A...564L...7K}}) is another object that was identified in 
the VLT-FLAMES Tarantula Survey (\protect{\cite{2015A&A...574A..13E}}) based on its blue emission-line 
spectrum. As with VFTS 1003, a red spectrum is needed to check for the presence of [O\,{\sc i}] 
emission. And alike that star, VFTS 822 has been proposed as possible Herbig B[e] pre-main sequence 
candidate. It displays a UV excess typical for pre-main sequence stars, but its luminosity of $\log 
L/L_{\odot} \sim 4$ is not conclusive and leaves as well room for an interpretation as evolved star. In 
the near-IR diagram VFTS 822 appears within the B[e]SG domain with a clear separation from the region 
hosting HAeBe stars, but from the WISE diagram the star is located at the border region with the HAeBes.

\paragraph{\bf ({\sl \textbf{VFTS 698}})}
This is the third object identified within in the VLT-FLAMES Tarantula Survey and suggested as B[e]SG 
(\protect{\cite{2012A&A...542A..50D}}). It also lacks a red optical spectrum to ascertain the presence 
of [O\,{\sc i}] emission in order to be considered a B[e] star. Its spectral variability and photometric 
light curve unveiled that the object is most likely an interacting binary consisting possibly of an 
early B-type star and a veiled more massive companion. Its IR colors place it in both diagrams into the 
region occupied by HAeBe stars, but the luminosity estimates for both stars with $\log L/L_{\odot} 
\sim 5$ and $5.3$ are too high for a pre-main sequence classification. Correcting for the 
extinction towards VFTS 698 would shift the object closer to the LBV domain. In the WISE diagram, the 
object appears particularly far off from the location of the classical B[e]SGs.  
Hence, this object needs further investigations for a proper classification.

\paragraph{\bf ({\sl \textbf{[L72] LH 85-10}})} 
Searching for LBV candidates in OB associations in the LMC the object number 10, within the 
association LH 85, was found to be a luminous emission-line star. Due to the similarity of its blue 
spectrum with other B[e]SGs, it has been suggested as a new member of that group 
(\protect{\cite{2000AJ....119.2214M}}). To date, no information about the presence of [O\,{\sc i}] 
emission from that object is available. Moreover, its spectral energy distribution does not display the 
characteristic near-IR excess emission of B[e]SGs questioning the validity of this classification 
(\cite{2009AJ....138.1003B}). The star's position in the near-IR diagram coincides with the region 
populated by LBVs, supporting the erroneous classification. No WISE photometry is available for this 
object.

\paragraph{\bf ({\sl \textbf{NOMAD1 0181-0125572}})} 

This star appears neither in the 2MASS nor in the WISE catalog, although it was reported to have 
2MASS photometry (\protect{\cite{2014A&A...568A..28L}}). From the coordinates of the source 
labeled NOMAD1 0181-0125572 by \cite{2011IAUS..272..260M} it is obvious, that it was confused with
the object LHA\,120-S\,165. The star with SIMBAD identifier NOMAD1 0181-0125572 is not a B[e]SG.


\subsubsection{Small Magellanic Cloud}

In the SMC, a total of six new B[e]SGs has been reported. Five of them were discovered as by-products 
from the spectroscopic survey of the hot, luminous stars in the SMC (2dFS, \cite{2004MNRAS.353..601E}), 
from the Runaways and Isolated O Star Spectroscopic Survey of the SMC (RIOTS4, 
\cite{2012ApJ...759...10G}), and from the SMC photometric catalog (\cite{2010AJ....140..416B}). One more 
object was serendipitously detected (\cite{2007ApJ...670.1331W}). From a critical inspection of the 
properties of these six stars, only one object appears to fulfill all required criteria, one object 
appears as very promising candidate that lacks only some complementary information for unambiguous 
classification, whereas the remaining four stars are considered as misclassified. All objects are 
presented in the following, their IR colors are provided in Table\,\ref{tab:SMC}, and their positions 
within the two color-color diagrams were included in Figure\,\ref{fig:color-color-MCnew}.

\paragraph{\bf [MA93] 1116 = NGC 346:KWBBe 200}
The star is one of the few objects residing in a cluster, the SMC cluster NGC 346. Due to its 
H$\alpha$ emission it was first classified as a compact H\,{\sc ii} region (\cite{1993A&AS..102..451M}), 
then as a Be star based on its optical photometry (\cite{1999A&AS..134..489K}), and in the following as 
B[e]SG based on its optical appearance and near-IR excess emission (\cite{2007ApJ...670.1331W}). 
In the mid-IR, [MA93] 1116 displays silicate emission and strong PAH bands, features often seen in 
HAeBe stars (e.g., \cite{2008ApJ...684..411K, 2002A&A...390.1089P}). Additional cold dust is 
surrounding the object as is implied by its detection at 24\,$\mu$m with Spitzer/MIPS. These 
characteristics led to the suggestion that [MA93] 1116 might be an evolved young stellar object 
(\cite{2015MNRAS.451.3504R}) or a HAeBe star (\cite{2013ApJ...771...16W}), despite of the lack of clear 
evidence for infall, e.g., in form of inverse P Cygni profiles. If true, its luminosity of $\log 
L/L_{\odot} \sim 4.4$ (\cite{2007ApJ...670.1331W, 2019MNRAS.488.1090C}) would place it at the upper 
limit for Herbig objects, while it would get in lane with the group of less-luminous B[e]SGs in the 
evolved scenario. The position of [MA93] 1116 within the WISE diagram leaves room for a possible 
classification as HAeBe star where it falls on the edge of the region populated by HAeBe stars, whereas 
it appears clearly off the HAeBe region in the near-IR color-color diagram. It is hence considered as 
B[e]SG, and the fifth object within a cluster (the first one in the SMC).

\begin{table}[H]
\caption{Confirmed and candidate B[e]SGs in the SMC. Misclassified objects are listed in the bottom part of the table.}
\centering
\begin{tabular}{lccccccc}
\toprule
\textbf{Object}	& \textbf{Ref.} & \sl{\textbf{E(B-V)}} & \textbf{Ref.} & \sl{\textbf{J-H}} & \sl{\textbf{H-K}}  & \sl{\textbf{W1-W2}} & \sl{\textbf{W2-W4}}  \\
\midrule
\multicolumn{8}{c}{Confirmed B[e]SGs}\\
\midrule
$[$MA93] 1116     &   \protect{\cite{2007ApJ...670.1331W, 2019MNRAS.488.1090C}} & 0.42 & \protect{\cite{2019MNRAS.488.1090C}} & 0.665$\pm$0.077 &  1.084$\pm$0.0689 & 1.112$\pm$0.03 &  6.883$\pm$0.034 \\
\midrule
\multicolumn{8}{c}{Uncertain or controversial classification}\\
\midrule
(LHA\,115-S\,38)     & \protect{\cite{2010AJ....140..416B}} & 0.13 & \protect{\cite{2014MNRAS.439.2211K}}&  0.468$\pm$0.038 & 0.879$\pm$0.038 & 1.034$\pm$0.031 & 3.978$\pm$0.035 \\
\midrule
\multicolumn{8}{c}{Erroneous classification}\\
\midrule
({\sl LHA\,115-N82})       & \protect{\cite{1990A&A...234..233H, 2004MNRAS.353..601E}} & 0.12 & \protect{\cite{1990A&A...234..233H}} &  0.488$\pm$0.040 & 1.141$\pm$0.040 & 1.133$\pm$0.031 & 2.723$\pm$0.038 \\
({\sl LHA\,115-S\,29}) & \protect{\cite{2012ApJ...759...10G}} & 0.05 & \protect{\cite{2012ApJ...759...10G}} &  0.089$\pm$0.030 & 0.180$\pm$0.032 & 0.346$\pm$0.030 & 2.479$\pm$0.020\\ 
({\sl LHA\,115-S\,46}) & \protect{\cite{2012ApJ...759...10G}} & 0.05 & \protect{\cite{2012ApJ...759...10G}} &  0.096$\pm$0.037 & 0.204$\pm$0.040 & 0.308$\pm$0.030 & 2.108$\pm$0.020 \\
({\sl LHA\,115-S\,62}) & \protect{\cite{2012ApJ...759...10G}} & 0.11 & \protect{\cite{2012ApJ...759...10G}} &  0.084$\pm$0.039 & 0.175$\pm$0.048 & 0.315$\pm$0.034 & 3.076$\pm$0.024 \\
\bottomrule
\multicolumn{8}{p{.9\textwidth}}{{\bf Note:} IR photometry for 
all objects is taken from the 2MASS point source catalog ($J, H, K$, \cite{2003yCat.2246....0C}) and from the WISE All-Sky Data Release ($W1, W2, W4$, \cite{2012yCat.2311....0C}).}\\
\label{tab:SMC}
\end{tabular}
\end{table}

\paragraph{\bf (LHA\,115-S\,38 = 2dFS 1804)}
This object was found from the spectroscopic survey of the hot, luminous stars in the SMC 
(\protect{\cite{2004MNRAS.353..601E}}). The blue spectrum displays numerous emission lines, in 
particular of [Fe\,{\sc ii}], but no information about [O\,{\sc i}] emission is available. The spectral 
energy distribution indicates a near-IR excess (\protect{\cite{2010AJ....140..416B}}). With a luminosity 
estimate of $\log L/L_{\odot} \simeq 4.1$ the star was also considered as a post-AGB object 
(\cite{2014MNRAS.439.2211K}), but its positions within the two color-color diagrams places the star 
clearly within the B[e]SG domain. As such, it is a strong candidate for another low-luminosity B[e]SG.

\paragraph{\bf ({\sl \textbf{LHA\,115-N82 = 2dFS 2837 = LIN 495}})}

This star displays all spectroscopic B[e]SG characteristics (\protect{\cite{1990A&A...234..233H, 
2004MNRAS.353..601E}}) and an IR excess emission (\protect{\cite{2010AJ....140..416B}}). The optical 
spectrum was reported to be composite (\protect{\cite{2004MNRAS.353..601E}}), and the detected 
photospheric lines display radial velocity variations, whereas the emission from the circumstellar 
matter appears to be stable (\protect{\cite{2019MNRAS.488.1090C}}). The IR colors of LHA\,115-N82 locate 
it clearly within the B[e]SG domains. Curiously, the $V$ and $I$ band light curves display a long-term 
brightening, which resembles LBV outbursts and is not common in B[e]SGs. On the other hand, it displays 
both [O\,{\sc i}] and [Ca\,{\sc ii}] emission which are typically not seen in LBVs. Since the star has 
too low luminosity ($\log L/L_{\odot} \simeq 3.8$) to be an LBV (or even a B[e]SG), it was recently 
assigned a classification as "LBV imposter" (\protect{\cite{2019MNRAS.488.1090C}}). This object clearly 
requires further investigations to pin down its status.

\paragraph{\bf ({\sl \textbf{LHA\,115-S\,29 = RMC\,15}}), ({\sl \textbf{LHA\,115-S\,46 = RMC\,38}}), ({\sl \textbf{LHA\,115-S\,62 = RMC\,48}})} 

The blue optical spectra of these three objects show emission-line features similar to B[e]SGs, and 
the luminosities derived for these objects, ranging from $\log L/L_{\odot} \simeq 4.4$ to 4.8, assigns 
them a supergiant status (\protect{\cite{2012ApJ...759...10G}}). However, it is currently not known 
whether these objects display [O\,{\sc i}] line emission, one of the defining characters of the B[e] 
phenomenon. Due to the lack of a pronounced near-IR excess emission, these three stars have been 
proposed to be dust-poor B[e]SGs (\protect{\cite{2012ApJ...759...10G}}). However, the presence of warm 
dust is another main classification criteria for a star to be considered as B[e]SG. It is not surprising 
that all three objects fall clearly outside the B[e]SG domains in the color-color diagrams. Instead, 
their positions coincide with the regions populated by LBVs, which implies that these three stars have 
dense winds and circumstellar ionized gas, which is exemplified by their emission-line spectra. These 
objects require further investigations to unveil their true nature.


\subsection{Local Group galaxies beyond the Magellanic Clouds}\label{sect:LocalGroup}

Moving further away, beyond the Magellanic Clouds, we may expect to find B[e]SG stars and candidates 
in those galaxies, in which star-formation is ongoing. For these galaxies, surveys such as the LGGS 
project mentioned in Section\,\ref{sect:method}, but also earlier surveys (e.g., 
\cite{1996AJ....112.1450C, 1999A&AS..140..309F}) provide indispensable information on the population of 
luminous, evolved massive stars, and provide the base for systematic investigations for unambiguous 
classification of these objects.

\begin{figure}[H]
\centering
\includegraphics[width=11.5cm]{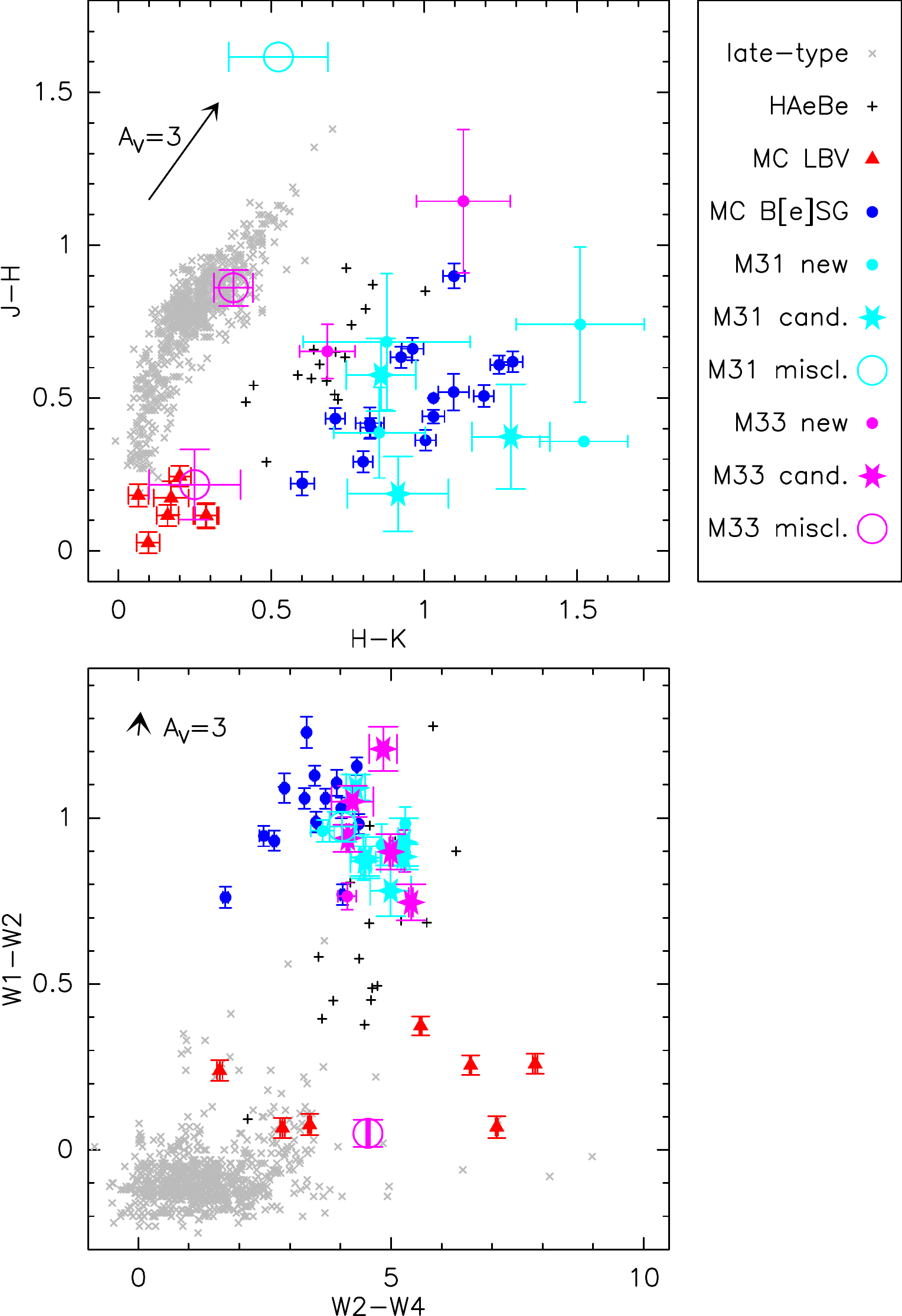}
\caption{As Figure\,\ref{fig:color-color-MCnew} but for the location of the M31 (light blue) and M33 
(purple) samples in the near-IR (top panel) and the WISE diagram (bottom panel) based on their observed 
colors. Typical values for the objects' reddening are $A_{\rm V} \leq 1.5$\,mag
(\cite{2017ApJ...844...40H}).}
\label{fig:JH-HK-M31-M33}
\end{figure}

To date, systematic spectroscopic studies have been performed in the two large spiral galaxies of 
the Local Group, M31 and M33, in which a number of B[e]SGs were found amongst the putative LBV 
candidates (\protect{\cite{2012A&A...541A.146C, 2014ApJ...780L..10K, 2015MNRAS.447.2459S, 
2017ApJ...836...64H, 2018MNRAS.480.3706K}}). The suggested B[e]SG populations in each of these 
galaxies is presented in the following subsections. The samples are provided in the 
Tables\,\ref{tab:M31} and \ref{tab:M33}, in which all objects are listed under their Local 
Group Galaxy Survey (LGGS) identifiers (column 1). The properties, based on which the decision to 
categorize an individual star as either a candidate or a misclassified object has been made, are 
briefly depicted. The locations of the confirmed, candidate, and possibly misclassified B[e]SGs in 
the two color-color diagrams are shown in Figure\,\ref{fig:JH-HK-M31-M33}.

\subsubsection{M31}

A total of 13 B[e]SGs has been proposed in M31 to date. Twelve of them resulted from optical 
spectroscopic observations of LBV candidates that display the typical B[e]SG characteristics 
(\cite{2017ApJ...836...64H, 2017ASPC..510..468S}). Infrared spectroscopic observations of a small LBV 
candidate sample revealed so far that three display CO band emission (\cite{2014ApJ...780L..10K,  
2015MNRAS.447.2459S}), making their classification as B[e]SG very likely, especially since two of the CO 
band emission objects were also found from optical spectroscopy to be a possible B[e]SG. 
However, a closer look at this total sample of 13 objects, including the available information 
from near- and mid-IR photometric observations, reveals that only four stars can be considered as 
confirmed B[e]SGs. Seven objects require further clarifications, and two appear to be misclassified 
(see Table\,\ref{tab:M31}).

\paragraph{\bf J004320.97+414039.6, J004415.00+420156.2, J004417.10+411928.0} 
These three objects fulfill all classification criteria of the confirmed B[e]SGs. Their
positions in the near-IR and WISE diagrams coincides with the domain populated by the B[e]SGs, except 
for J004320.97+414039.6, which appears a bit off due to the uncertainties in its $J$ and $H$ magnitudes, 
which are only upper limits.

\paragraph{\bf J004522.58+415034.8}  
The star was assigned a warm hypergiant status due to detected photospheric features pointing towards 
an A2Ia spectral-type (\cite{2013ApJ...773...46H}). As it otherwise fulfills all criteria of a B[e]SG 
(even displaying CO band emission, \protect{\cite{2014ApJ...780L..10K}}), it is included into the list 
of confirmed B[e]SGs, but keeping in mind that with its A spectral type it is actually a representative 
of the (though few) A[e]SGs. Recent modeling of the star's spectral energy distribution resulted in an 
effective temperature estimate of 11\,000\,K (\cite{2017ApJ...844...40H}), which would point towards a 
late-B spectral type.

\paragraph{\bf (J004220.31+405123.2), (J004221.78+410013.4), (J004229.87+410551.8), (J004442.28+415823.1)} 
These stars display the optical characteristics of B[e]SGs (\protect{\cite{2017ApJ...836...64H}}), and 
their WISE colors place them within the B[e]SG domain. But all four objects lack $JHK$ band photometry, 
so that they cannot be located in the $J-H$ versus $H-K$ diagram. The presence of He\,{\sc i} emission 
in J004221.78+410013.4 (\protect{\cite{2017ApJ...836...64H}}) contradicts the proposed effective 
temperature of 7200\,K (\cite{2017ApJ...844...40H}) and requires clarification.

\paragraph{\bf (J004411.36+413257.2)} The star is reported to display He\,{\sc i}, Fe\,{\sc ii}, and 
[Fe\,{\sc ii}] emission and was suggested as B[e]SG star (\protect{\cite{2017ASPC..510..468S}}). Whether 
it displays [O\,{\sc i}] emission was not discussed by these authors and is hence yet unclear, although 
the spectrum covers that spectral region. Its $J-H$ color
is smaller than for typical B[e]SGs (but the $H$ magnitude is uncertain), placing the star at the 
lower boundary for the B[e]SGs in the color-color diagram. No WISE photometry exists. The closest
IR source is more than 6 arcsec away. The object is not reported in other published spectroscopic 
investigations to display all B[e]SG characteristics. Instead, it was listed as Fe\,{\sc ii} emission 
line star (\cite{2014ApJ...790...48H}) and more recently as LBV candidate (\cite{2017ApJ...844...40H}).

\begin{table}[H]
\caption{Confirmed and candidate B[e]SGs in M31. Misclassified objects are listed at the bottom of the table.}
\centering
\begin{tabular}{lccccc}
\toprule
\textbf{LGGS }	&  \textbf{Ref.} & \sl{\textbf{J-H}} & \sl{\textbf{H-K}}  & \sl{\textbf{W1-W2}} & \sl{\textbf{W2-W4}} \\
\midrule
\multicolumn{6}{c}{Confirmed B[e]SGs}\\
\midrule
J004320.97+414039.6 & \protect{\cite{2017ApJ...836...64H, 2017ASPC..510..468S}} & {\sl 0.358}    & {\sl 1.522$\pm$0.143} & 0.983$\pm$0.051 & 5.284$\pm$0.084 \\
J004415.00+420156.2$^{a}$ & \protect{\cite{2017ApJ...836...64H, 2017ASPC..510..468S}} & 0.741$\pm$0.253 & 1.510$\pm$0.210 & 0.961$\pm$0.033 & 3.645$\pm$0.230 \\
J004417.10+411928.0$^{b,c}$ & \protect{\cite{2014ApJ...780L..10K, 2017ApJ...836...64H}} & 0.387$\pm$0.148 & 0.853$\pm$0.149 & 0.882$\pm$0.068 & 4.455$\pm$0.247 \\
J004522.58+415034.8$^{c,d}$ & \protect{\cite{2014ApJ...780L..10K}} & {\sl 0.684$\pm$0.223} & 0.877$\pm$0.273 & 1.091$\pm$0.039 & 4.303$\pm$0.187 \\
\midrule
\multicolumn{6}{c}{Uncertain or controversial classifications}\\
\midrule
(J004220.31+405123.2) & \protect{\cite{2017ApJ...836...64H}} &  \ldots         &    \ldots       & 0.880$\pm$0.062 & 4.495$\pm$0.298 \\ 
(J004221.78+410013.4) & \protect{\cite{2017ApJ...836...64H}} &  \ldots         &    \ldots       & 0.927$\pm$0.072 & 5.253$\pm$0.107 \\
(J004229.87+410551.8)$^{e}$ & \protect{\cite{2017ApJ...836...64H, 2017ASPC..510..468S}} &  \ldots         &    \ldots       & 0.870$\pm$0.045 & 4.484$\pm$0.082 \\
(J004411.36+413257.2) & \protect{\cite{2017ASPC..510..468S}} & {\sl 0.187$\pm$0.123}  & {\sl 0.914$\pm$0.165}  &  \ldots  & \ldots  \\
(J004442.28+415823.1) & \protect{\cite{2017ApJ...836...64H, 2017ASPC..510..468S}} &   \ldots        &                  
  \ldots        & 0.781$\pm$0.076 & 4.994$\pm$0.406\\
(J004444.52+412804.0)$^{c,d}$ & \protect{\cite{2015MNRAS.447.2459S}} & 0.576$\pm$0.119 & 0.858$\pm$0.114 & 0.883$\pm$0.039 &  5.247$\pm$0.046\\
(J004621.08+421308.2)$^{d}$  & \protect{\cite{2017ASPC..510..468S}} & {\sl 0.374$\pm$0.171} & {\sl 1.283$\pm$0.128} & 1.091$\pm$0.039 & 4.303$\pm$0.188 \\ 
\midrule
\multicolumn{6}{c}{Erroneous classifications}\\
\midrule
({\sl J004043.10+410846.0}) & \protect{\cite{2017ApJ...836...64H, 2017ASPC..510..468S}} & {\sl 1.616}     &  
{\sl 0.523$\pm$0.162} & 0.974$\pm$0.045  & 4.030$\pm$0.250 \\
({\sl J004057.03+405238.6}) & \protect{\cite{2017ApJ...836...64H}} & \ldots   &  \ldots  &  \ldots  & \ldots  \\ 
\bottomrule
\multicolumn{6}{p{.9\textwidth}}{{\bf Note:} IR photometry is taken from the 2MASS point source 
catalog ($J, H, K$, \cite{2003yCat.2246....0C}) and from the WISE All-Sky Data Release ($W1, W2, W4$, \cite{2012yCat.2311....0C}). Colors resulting from uncertain photometric values are written in italic.}\\
\multicolumn{6}{p{.9\textwidth}}{$^{a}$ Possible contamination in the photometric bands {\sl J, W1} and {\sl W2} due to crowding.}\\
\multicolumn{6}{p{.9\textwidth}}{$^{b}$ Possible contamination in the photometric bands {\sl J} and {\sl H} due to crowding.}\\
\multicolumn{6}{p{.9\textwidth}}{$^{c}$ Has CO band emission (\protect{\cite{2014ApJ...780L..10K, 2015MNRAS.447.2459S}}}\\
\multicolumn{6}{p{.9\textwidth}}{$^{d}$ Also classified as warm hypergiant (\cite{2013ApJ...773...46H}).}\\
\multicolumn{6}{p{.9\textwidth}}{$^{e}$ Possible contamination in the photometric bands {\sl W1} and {\sl W2} due to crowding.}\\
\label{tab:M31}
\end{tabular}
\end{table}

\paragraph{\bf (J004444.52+412804.0)} 
This object was first proposed to be a P Cyg-type LBV candidate (\cite{2007AJ....134.2474M}). With the 
detection of [Ca\,{\sc ii}] emission and circumstellar dust (\cite{2014ApJ...790...48H}), and with the 
determination of an effective temperature of 6600\,K (\cite{2017ApJ...844...40H}), it was classified as 
a warm hypergiant. On the other hand, a B[e]SG classification with a central star of 15\,000--20\,000\,K 
has been suggested, based on the presence of He\,{\sc i} lines displaying P Cyg profiles and the 
detection of intense CO band emission (\protect{\cite{2015MNRAS.447.2459S}}). This controversial 
classification, based on spectra taken at similar epochs, requires clarification. Infrared photometry 
places J004444.52+412804.0 within the region occupied by B[e]SGs in the two color-color diagrams.

\paragraph{\bf (J004621.08+421308.2)}  
A status of a late A-type warm hypergiant has been assigned to this object based on the lack of 
detected He\,{\sc i} emission (\cite{2016ApJ...825...50G}).  In contrast to this,  He\,{\sc i} emission
lines were reported by \protect{\cite{2017ASPC..510..468S}}, and an effective temperature of 10\,000\,K 
has been estimated from the modeling of the stars' spectral energy distribution 
(\cite{2017ApJ...844...40H}). The position of the object in the near-IR and WISE diagrams coincides with 
the B[e]SG domain.

\paragraph{\bf ({\sl \textbf{J004043.10+410846.0}})} 
Its optical spectrum displays all typical characteristics of a B[e]SG 
(\protect{\cite{2017ApJ...836...64H}}), and its WISE colors confirm the presence of warm dust, but its 
(though uncertain) unusually high $J-H$ value of 1.6 at a low value of $H-K$ fits neither to B[e]SGs nor 
to LBVs.

\paragraph{\bf ({\sl \textbf{J004057.03+405238.6}})} 
No IR photometry is available, so that currently no statement can be made about the presence of (hot) 
dust. The absence of He\,{\sc i} lines in the optical spectra along with an effective temperature of 
7700\,K (\cite{2017ApJ...844...40H}) speak against a B[e]SG classification.

\subsubsection{M33}

Also in M33, the sample of putative LBV candidates (\cite{2007AJ....134.2474M, 2016AJ....152...62M}) 
served as a starting point for a more detailed investigation of the individual objects. Optical 
(\cite{2005A&A...437..217F, 2012A&A...541A.146C, 2017ApJ...836...64H}), and infrared 
(\cite{2018MNRAS.480.3706K}) spectroscopic studies revealed so far a total of ten possible B[e]SGs. They 
are listed in Table\,\ref{tab:M33}. Combined with the available information about their infrared colors, 
only two objects can currently be considered as confirmed B[e]SGs, six need further clarification, and 
two objects appear to be misclassified.

\paragraph{\bf J013333.22+303343.4, J013350.12+304126.6}
Both objects display all classification characteristics of B[e]SGs. Their locations in the WISE 
diagram coincide with the B[e]SG domain. In the near-IR diagram they appear slightly off the B[e]SG 
region, shifted towards the HAeBe domain, but their high-luminosities clearly classify them as 
supergiants. The CO band absorption detected from J013333.22+303343.4 cannot be explained with a 
cool companion, but might originate from a pole-on seen jet (\cite{2018MNRAS.480.3706K}). Whether also 
CO band emission is present in this object, but veiled by the intense absorption component, is currently 
unknown. 

\paragraph{\bf (J013324.62+302328.4), (J013342.78+303256.3), (J013349.28+305250.2), (J013426.11+303424.7), (J013459.47+303701.9), (J013500.30+304150.9)} 
All six objects lack near-IR photometry, which renders it difficult to unambiguously classify them 
as B[e]SGs, although their mid-IR colors and their optical spectroscopic appearances support such a 
classification. The star J013342.78+303256.3 has even no mid-IR photometry, so that it is unclear 
whether it is surrounded by (warm) dust at all. There is an IR source at a distance of $\sim$ 4.5 arcsec 
which is considered as too far off to be identified with J013342.78+303256.3. The sources 
J013324.62+302328.4 and J013500.30+304150.9 were reported as intrinsically faint objects with little 
extinction, and were suggested as counterparts of the low-luminosity B[e]SG in the MCs 
(\cite{2012A&A...541A.146C}). But according to the luminosity estimates for these stars, for which 
values of $\log L/L_{\odot} \simeq 4.8-5.0$ were obtained (\cite{2017ApJ...844...40H}), these stars fall 
rather onto the lower boundary of the high-luminosity B[e]SGs sample within the MCs (see 
Figure\,\ref{fig:HRD-classic}).

\paragraph{\bf ({\sl \textbf{J013242.26+302114.1}})} The optical spectra of this object display
all typical B[e]SG characteristics, but the star's IR colors displace it clearly from the sample of 
confirmed B[e]SGs. It has excess emission at 8\,$\mu$m due to PAHs (\cite{2017ApJ...836...64H}). 
The $JHK$ magnitudes suggest that J013242.26+302114.1 falls into the region of 
late-type stars, respectively stars with a late-type companion (\cite{1985A&AS...61..237S, 
1995A&A...302..409G}). This classification is supported by the WISE colors, although regular late-type 
stars typically have slightly negative $W1-W2$ colors. Considering only the WISE photometry, 
J013242.26+302114.1 would fit to the region populated by LBVs. This objects clearly needs to be 
investigated in more detail for a proper classification.

\paragraph{\bf ({\sl \textbf{J013406.63+304147.8}})} This star is also known as [HS80] B416. Despite 
the detected CO band emission from this object (\cite{2018MNRAS.480.3706K}), a classification as B[e]SG 
seems not appropriate (as previously mentioned by \cite{2014ApJ...790...48H, 2017ApJ...836...64H}), 
because its near-IR colors show clearly the lack of warm dust and would place 
the object to the region populated by LBVs. Also no [O{\sc i}] emission is seen from 
J013406.63+304147.8 (\cite{2014ApJ...790...48H}). But so far no photometric variability, typical for 
LBVs, is recorded (\cite{2012A&A...541A.146C}), and the star can only be considered as a possible LBV 
candidate, as was previously suggested (\cite{2017ApJ...836...64H}). The star is found to be surrounded 
by an expanding ring-like nebula (\cite{2005A&A...437..217F}) and shows excess emission at 8\,$\mu$m due 
to PAHs (\cite{2018MNRAS.480.3706K}). The spectral energy distribution leaves space for a companion 
(\cite{2014ApJ...790...48H, 2018MNRAS.480.3706K}), and its spectral lines display radial velocity 
variations with a period of 16.13\,d, leading to the suggestion that the system might be an interacting 
binary system causing mass loss in the equatorial plane (\cite{2004BaltA..13..156S}).

\begin{table}[H]
\caption{Confirmed and candidate B[e]SGs in M33. Misclassified objects are listed at the bottom of the table.}
\centering
\begin{tabular}{lccccc}
\toprule
\textbf{LGGS }	&  \textbf{Ref.} & \sl{\textbf{J-H}} & \sl{\textbf{H-K}}  & \sl{\textbf{W1-W2}} & \sl{\textbf{W2-W4}} \\
\midrule
\multicolumn{6}{c}{Confirmed B[e]SGs}\\
\midrule
J013333.22+303343.4$^{a}$ & \protect{\cite{2012A&A...541A.146C, 2017ApJ...836...64H, 2018MNRAS.480.3706K}} & 1.144$\pm$0.234 & {\sl 1.128$\pm$0.154} & 0.900$\pm$0.063 & 5.269$\pm$0.098 \\
J013350.12+304126.6 & \protect{\cite{2012A&A...541A.146C, 2017ApJ...836...64H}} & 0.653$\pm$0.089 & 0.683$\pm$0.091 &  0.764$\pm$0.041 & 4.132$\pm$0.186 \\
\midrule
\multicolumn{6}{c}{Uncertain or controversial classifications}\\
\midrule
(J013324.62+302328.4) & \protect{\cite{2012A&A...541A.146C, 2017ApJ...836...64H}} & \ldots & \ldots & 1.050$\pm$0.047 & 4.235$\pm$0.414 \\
(J013342.78+303256.3) & \protect{\cite{2017ApJ...836...64H}} & \ldots & \ldots & \ldots & \ldots \\
(J013349.28+305250.2) & \protect{\cite{2017ApJ...836...64H}} & \ldots & \ldots & 0.746$\pm$0.054 & {\sl 5.394$\pm$0.040}\\
(J013426.11+303424.7) & \protect{\cite{2012A&A...541A.146C, 2017ApJ...836...64H}} & \ldots & \ldots & 1.208$\pm$0.066 & 4.845$\pm$0.278 \\
(J013459.47+303701.9) & \protect{\cite{2012A&A...541A.146C, 2017ApJ...836...64H}} & \ldots & \ldots & 0.898$\pm$0.053 & {\sl 4.996$\pm$0.038} \\
(J013500.30+304150.9) & \protect{\cite{2012A&A...541A.146C, 2017ApJ...836...64H}} & \ldots & \ldots & 0.939$\pm$0.040 & {\sl 4.142$\pm$0.029}\\
\midrule
\multicolumn{6}{c}{Erroneous classifications}\\
\midrule
({\sl J013242.26+302114.1}) & \protect{\cite{2012A&A...541A.146C, 2017ApJ...836...64H}} &  0.861$\pm$0.059 & 0.376$\pm$0.064 & 0.050$\pm$0.040 & {\sl 4.543$\pm$0.031} \\
({\sl J013406.63+304147.8})$^{b}$ & \protect{\cite{2005A&A...437..217F, 2012A&A...541A.146C, 2018MNRAS.480.3706K}}  & 0.217$\pm$0.115 & 0.249$\pm$0.149 & \ldots & \ldots \\
\bottomrule
\multicolumn{6}{p{.9\textwidth}}{{\bf Note:} IR photometry is taken from the 2MASS point source catalog ($J, H, K$, \cite{2003yCat.2246....0C}) and from the WISE All-Sky Data Release ($W1, W2, W4$,\cite{2012yCat.2311....0C}). Colors resulting from uncertain photometric values are written in italic.}\\
\multicolumn{6}{p{.9\textwidth}}{$^{a}$ Has CO band absorption (\cite{2018MNRAS.480.3706K}).}\\
\multicolumn{6}{p{.9\textwidth}}{$^{b}$ Has CO band emission (\cite{2018MNRAS.480.3706K}).}\\
\label{tab:M33}
\end{tabular}
\end{table}


\subsection{Byond the Local Group}\label{sect:BeyondLG}
    
Not much is known yet about the B[e]SG population beyond the Local Group. Most promising for dedicated 
searches are the nearby large spiral galaxies M81, M101, and NGC 2403, for which pioneering
ground-based surveys have been conducted already in the eighties and nineties of the last century 
(\cite{1983AJ.....88.1569S, 1984AJ.....89..621S, 1984AJ.....89..630S, 1991AJ....102..113Z}) and which 
are nowadays extended to fainter objects thanks to the capabilities of the {\sl Hubble Space Telescope} 
(e.g., \cite{2013AJ....146..114G}). Spectroscopic follow-up investigations revealed a number of variable
luminous objects (\cite{1980ApJ...241..598H, 1987AJ.....94.1156H, 1998AstL...24..507S, 
1998AstL...24..603S, 2015AJ....149..152G}), but only in one of these galaxies, M81, three B[e]SG 
candidates have been found so far (\cite{2019AJ....157...22H}). They all display the typical B[e] 
features in their optical spectra, but without complementary information about the presence of (warm) 
circumstellar dust, their classification remains preliminary. These objects will certainly not remain 
the only of their kind, because the search for more candidates has just begun.

The B[e]SG candidates in M81 are listed in Table\,\ref{tab:M81}. For lack of proper SIMBAD identifiers 
for these objects, the table contains the star ID (\protect{\cite{2019AJ....157...22H}}) along with the 
coordinates and, where available, the $V$ band magnitudes.

\begin{table}[H]
\caption{Candidate B[e]SGs in M81}
\centering
\begin{tabular}{lcccc}
\toprule
\textbf{Star ID} & \textbf{RA(J2000)} & \textbf{Dec(J2000)} & \sl{\textbf{V}} &  \textbf{Ref.} \\
\midrule
(10584-8-4)      &  9:54:50.03        & +69:06:55.47        & \ldots & \protect{\cite{2019AJ....157...22H}}\\
(10584-4-1)      &  9:54:54.05        & +69:10:23.00        &  19.68 & \protect{\cite{2019AJ....157...22H}}\\
(10584-9-1)      &  9:55:18.97        & +69:08:27.54        &  19.10 & \protect{\cite{2019AJ....157...22H}}\\
\bottomrule
\label{tab:M81}
\end{tabular}
\end{table}

In the past few years, additional surveys have been performed, using large ground-based facilities 
(e.g., \cite{2017MNRAS.465.4985P}) and space telescopes (e.g., \cite{2018A&A...618A.185S, 
2019arXiv190707140S}). These surveys were aimed at revealing the luminous and variable massive star 
populations of other galaxies even further away. They provide an excellent basis for follow-up 
spectroscopic studies to classify their massive star content, so that many more LBV candidates and 
B[e]SGs may be found in the (near) future. Also the Local Group galaxies still need to be explored in 
more detail. In particular those galaxies, that were already found to possess (even though in very small 
numbers) LBV candidates (\cite{2007AJ....134.2474M}) that are awaiting their proper classifications. The 
upcoming era of the Extreme Large Telescope (ELT) promises to become particularly fruitful. The next 
generation of high-sensitivity instruments combined with the large collecting area of the telescope will 
facilitate ground-based spectroscopic observations of faint objects with very high spatial resolution.


\subsection{Galactic objects}\label{sect:MilkyWay}

Finally, we return to our own Galaxy. Searching for B[e]SG stars in the Milky Way is a difficult task. 
Though many B[e] stars are known, the assignment of a supergiant status is significantly hampered due 
to highly uncertain distances hence luminosities. The situation will hopefully change with the final 
data release from the GAIA mission, from which one hopes for accurate parallax measurements. But for 
now, the luminosities of the objects are subject to large uncertainties, so that only objects with a 
reported luminosity of at least $\log L/L_{\odot} \ge 5$ are considered as serious B[e]SG candidates. 

The lower luminosity boundary of $\log L/L_{\odot} \sim 4.0$ for an evolved star to be assigned a 
supergiant status is a further hindrance in the classification of objects as B[e]SGs, because 
this luminosity domain is shared with the massive pre-main sequence (HAeBe) 
stars. The latter have emission-line spectra with numerous forbidden emission lines from [Fe\,{\sc ii}] 
and [O\,{\sc i}] alike the B[e]SGs, and the stars are surrounded by significant amounts of circumstellar 
dust within their massive accretion disks causing considerable IR excess emission, just as the B[e]SGs.
Hence, it is not surprising that confusion exists about the proper classification for a number of objects 
within this luminosity domain of $4.0 < \log L/L_{\odot} < 4.5$, and that Galactic B[e]SG candidates 
also appear as candidates in catalogs of HAeBe stars (see, e.g., \cite{2017MNRAS.472..854A}). In the 
absence of clear indications for infall of material, which is a typical characteristic of pre-main 
sequence stars, alternative discriminators for the classification of such objects are needed. 

A reasonable approach to this is to search for $^{13}$CO emission from the circumstellar environments 
of the uncertain candidates. Many HAeBe stars have been reported to display CO band emission from 
their massive accretion disks (e.g., \cite{2004A&A...427L..13B, 2004ApJ...617.1167B, 
2013MNRAS.429.2960I, 2014MNRAS.445.3723I, 2018MNRAS.477.3360I}). 
As these disks form from material provided by the interstellar medium in which the $^{12}$C/$^{13}$C 
isotope abundance ratio has typically a value of about 90 (\cite{2012A&A...537A.146E}), these
pre-main sequence disks can clearly be distinguished from the disks around evolved massive stars,
which should be enriched in $^{13}$C and hence give rise to clearly measurable emission in $^{13}$CO. 
$K$-band spectra of these objects, covering the first-overtone bands of both $^{12}$CO and 
$^{13}$CO (see Section\,\ref{sect:current_evol} and Figure\,\ref{fig:13CO}), are thus key for a 
proper discrimination between a young (pre-main sequence) and an evolved status. 

From the currently proposed 15 Galactic B[e]SGs listed in Table\,\ref{tab:CO-O-Ca}, nine have been 
reported to display CO band emission. But so far, only four of them have been observed in the region 
around the $^{13}$CO bands. The spectra of all four stars have been found to display clear signatures of 
$^{13}$CO emission, and model results revealed that the environments of all four objects are clearly 
enriched in $^{13}$C. Two of these objects were already known to be supergiants based on their confirmed 
high luminosities. These are the stars GG\,Car (\protect{\cite{2013A&A...549A..28K, 
2013A&A...558A..17O}}) and Hen 2-398 (\protect{\cite{2013A&A...558A..17O}}). For the other two, which so 
far have also been considered as HAeBe candidates (see Table\,\ref{tab:Galactic}), the detection of 
chemically processed material can hence be regarded as the ultimate proof of their evolved, supergiant 
nature. These are the objects MWC\,137 (\cite{2015AJ....149...13M}) and MWC\,349 
(\cite{Kraus2019_prep}). These results are very promising and encouraging, and they demonstrate that the 
$^{13}$CO molecular emission provides a solid tool to unambiguously classify a star as either a pre-main 
sequence or an evolved object. Clearly more observational effort needs to be undertaken to search also 
for the signatures of $^{13}$CO in the spectra of the remaining objects.

\begin{figure}[H]
\centering
\includegraphics[width=11.5cm]{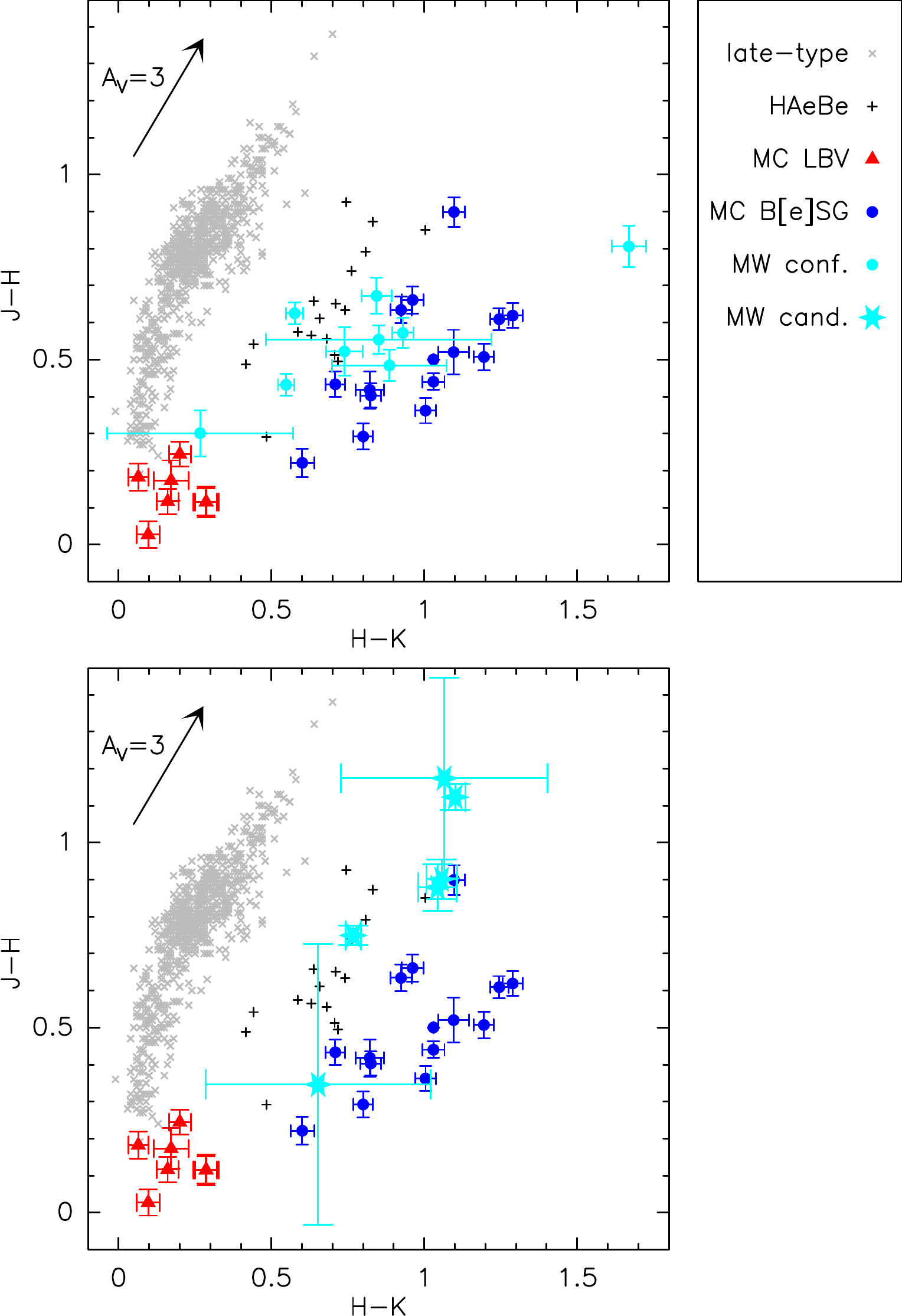}
\caption{Near-IR diagrams as in Figure\,\ref{fig:color-color-MCnew}, showing the locations of the 
Galactic confirmed B[e]SGs (top panel) and B[e]SG candidates (bottom panel). The colors of the Galactic 
objects have been corrected for interstellar extinction (Table\,\ref{tab:Galactic}).}
\label{fig:Galactic}
\end{figure}

When collecting the IR magnitudes of the Galactic sample, it turned out that only the near-IR 
measurements are reliable, whereas the WISE measurements for all objects have been flagged as being 
contaminated by neighboring objects. The latter are hence useless for classification purposes, and one 
can currently only rely on the $JHK$-band magnitudes. The list of objects, their observed colors, and
literature values of their color excess are listed in Table\,\ref{tab:Galactic}. The relatively high
values of the color excess requires correction for extinction before placing the objects to the near-IR
diagram. Corrections have been performed with the galactic extinction curve using
an $R_{\rm V}$  value of 3.1 has been utilized (\cite{1989ApJ...345..245C}). The extinction corrected
colors are included in Table\,\ref{tab:Galactic}, and the positions of the objects are shown in 
Fig.\,\ref{fig:Galactic} separately for the confirmed (left panel) and candidate objects (right panel).
 
\begin{table}[H]
\caption{Confirmed and candidate B[e]SGs in the Milky Way.}
\centering
\begin{tabular}{lccccccc}
\toprule
\textbf{Object}	&  \textbf{Ref.} & \sl{\textbf{E(B-V)}} & \textbf{Ref.} & \sl{\textbf{(J-H)}} & \sl{\textbf{(H-K)}}  & \sl{\textbf{(J-H)$_{0}$}} & \sl{\textbf{(H-K)$_{0}$}} \\
\midrule
\multicolumn{8}{c}{Confirmed B[e]SGs}\\
\midrule
MWC\,137$^{a}$    & \protect{\cite{1998MNRAS.298..185E, 2015AJ....149...13M}}                      &  1.22     & \protect{\cite{1998MNRAS.298..185E}}  &  0.922$\pm$0.040 & 1.217$\pm$0.035 & 0.572$\pm$0.040 & 0.930$\pm$0.035  \\
MWC\,349$^{a}$    & \protect{\cite{2012A&A...541A...7G, Kraus2019_prep}}                           & $\sim$ 3.2& \protect{\cite{1985ApJ...292..249C}}  &  1.472$\pm$0.038 & 1.603$\pm$0.369 & 0.554$\pm$0.038 & 0.851$\pm$0.369  \\
GG\,Car$^{b}$     & \protect{\cite{2012A&A...540A..91M, 2013A&A...549A..28K, 2013A&A...558A..17O}} &  0.51     & \protect{\cite{2012A&A...540A..91M}}  &  0.818$\pm$0.049 & 0.964$\pm$0.049 & 0.672$\pm$0.049 & 0.844$\pm$0.049  \\
Hen 3-298         & \protect{\cite{2005A&A...436..653M, 2013A&A...558A..17O}}                      & 1.7       & \protect{\cite{2005A&A...436..653M}}  &  1.009$\pm$0.065 & 1.139$\pm$0.060 & 0.522$\pm$0.065 & 0.739$\pm$0.060  \\
CPD-52 9243       & \protect{\cite{2012A&A...548A..72C}}                                           & 1.7       & \protect{\cite{1988ApJ...324.1071M}}  &  0.919$\pm$0.029 & 0.948$\pm$0.026 & 0.432$\pm$0.029 & 0.548$\pm$0.026  \\
HD\,327083$^{b}$  & \protect{\cite{2003A&A...406..673M}}                                           & 1.8       & \protect{\cite{2003A&A...406..673M}}  &  1.000$\pm$0.042 & 1.309$\pm$0.188 & 0.484$\pm$0.042 & 0.886$\pm$0.188  \\
MWC\,300$^{a,b}$  & \protect{\cite{1985A&A...148..412W, 2004A&A...417..731M}}                      & 1.2       & \protect{\cite{2004A&A...417..731M}}  &  1.150$\pm$0.056 & 1.951$\pm$0.055 & 0.806$\pm$0.056 & 1.669$\pm$0.055  \\
AS 381$^{b}$      & \protect{\cite{2002A&A...383..171M}}                                           & 2.3       & \protect{\cite{2002A&A...383..171M}}  &  1.285$\pm$0.029 & 1.117$\pm$0.028 & 0.625$\pm$0.029 & 0.576$\pm$0.028  \\
CPD-57 2874       & \protect{\cite{2007A&A...464...81D}}                                           & 1.75      & \protect{\cite{2007A&A...464...81D}}  &  0.803$\pm$0.062 & 0.678$\pm$0.304 & 0.301$\pm$0.062 & 0.267$\pm$0.304  \\
\midrule
\multicolumn{8}{c}{Uncertain or controversial classifications}\\
\midrule
(Hen 3-938)         & \protect{\cite{2019MNRAS.488.1090C}}                       & 1.64      & \protect{\cite{2019MNRAS.488.1090C}}                      &  1.593$\pm$0.035 & 1.487$\pm$0.034 & 1.124$\pm$0.035 & 1.101$\pm$0.034  \\
(MWC\,342)$^{a,b}$  & \protect{\cite{1999A&A...349..126M}}                       & 1.5       & \protect{\cite{1999A&A...349..126M}}                      &  1.179$\pm$0.027 & 1.121$\pm$0.025 & 0.749$\pm$0.027 & 0.768$\pm$0.025  \\
(Hen 3-303)         & \protect{\cite{2005A&A...436..653M}}                       & 1.7       & \protect{\cite{2005A&A...436..653M}}                      &  1.366$\pm$0.063 & 1.443$\pm$0.063 & 0.879$\pm$0.063 & 1.042$\pm$0.063  \\
(CD-42 11721)$^{a}$ & \protect{\cite{2007MNRAS.377.1343B}}                       & 1.4-1.6   & \protect{\cite{2012A&A...548A..72C, 2007MNRAS.377.1343B}} &  1.317$\pm$0.053 & 1.398$\pm$0.050 & 0.901$\pm$0.053 & 1.057$\pm$0.050  \\
(HD\,87643)$^{a,b}$ & \protect{\cite{1988ApJ...324.1071M, 2009A&A...507..317M}}  & 1.0       & \protect{\cite{1988ApJ...324.1071M}}                      &  1.461$\pm$0.271 & 1.300$\pm$0.338 & 1.174$\pm$0.271 & 1.065$\pm$0.338  \\
(HD\,62623)$^{b}$   & \protect{\cite{2010AstBu..65..150C}}                       & 0.17      & \protect{\cite{1995A&A...293..363P}}                      &  0.395$\pm$0.380 & 0.693$\pm$0.368 & 0.346$\pm$0.380 & 0.653$\pm$0.368  \\
\bottomrule
\multicolumn{8}{p{.95\textwidth}}{{\bf Note:} IR photometry is taken from the 2MASS point source catalog ($J, H, K$, \cite{2003yCat.2246....0C}).}\\
\multicolumn{8}{p{.95\textwidth}}{$^{a}$ Star appears also in HAeBe catalogs (see, e.g., \protect{\cite{2017MNRAS.472..854A}}).}\\
\multicolumn{8}{p{.95\textwidth}}{$^{b}$ Confirmed or suspected binary.}\\
\label{tab:Galactic}
\end{tabular}

\end{table}

The separation of confirmed from candidate B[e]SGs is based on two characteristics: (i) Stars
with detected enrichment in $^{13}$CO of their circumstellar environments are considered as confirmed, 
as well as (ii) stars with reported (by more than one research team) luminosity values of $\log 
L/L_{\odot} \ge 5.0$. Objects with lower luminosities $4.0 < \log L/L_{\odot} < 5.0$ are assigned 
a candidate status. Based on these criteria, the Galactic sample splits into nine confirmed B[e]SGs and 
six candidates (see Table\,\ref{tab:Galactic}). 

\paragraph{\bf MWC\,137, MWC\,349, GG\,Car, Hen\,3-298}
These four objects are considered as confirmed B[e]SGs based on the detected enrichment of their
circumstellar disk material with $^{13}$CO (see Table\,\ref{tab:CO-O-Ca}). They all fall into the 
region of the confirmed B[e]SGs in the near-IR diagram, regardless of the large errorbar for MWC\,349. 

\paragraph{\bf CPD-52 9243, HD\,327083, MWC\,300}
All three stars fulfill the high luminosity criterion. The near-IR colors of CPD-52 9243 and HD\,327083 
place these two objects within the B[e]SG domain. MWC\,300 appears regularly in studies of 
HAeBe stars. Its near-IR colors locate this star close to the B[e]SGs but far away from the HAeBe 
region, making a pre-main sequence nature of this object rather unlikely. Its relatively high color 
values might be influenced by a possible companion (\protect{\cite{2004A&A...417..731M, 
2012A&A...545L..10W}})

\paragraph{\bf AS\,381, CPD-57\,2874}
These two objects are also known to have high luminosities, but they reside slightly outside the 
classical B[e]SG domain in the near-IR diagram. AS\,381 is a reported binary 
(\protect{\cite{2002A&A...383..171M, 2014MNRAS.443..947L}}) consisting of a luminous B[e]SG and a K-type 
companion, which seems to (significantly) contribute to the total near-IR flux, hence altering the colors 
of the B[e]SG. The near-IR colors of CPD-57\,2874 place it closer to the LBVs rather than to the B[e]SGs, 
although its $H-K$ color is subject to large uncertainty. Its pronounced emission in CO, and in the 
[O\,{\sc i}] and [Ca\,{\sc ii}] forbidden lines (\protect{\cite{2018MNRAS.480..320M}}) speak against an 
LBV classification.

\paragraph{\bf (Hen\,3-938), (MWC\,342), (Hen\,3-303), (CD-42\,11721)}

The rather low luminosities of these four objects and the closeness or even coincidence of their
location with the HAeBe domain requires clearly further studies for an unambiguous classification.
While Hen\,3-938 and MWC\,342 have to our knowledge not yet been spectroscopically observed in the 
$K$-band, the other two stars showed no evidence for CO band emission (\cite{2012BAAA...55..123M}). This 
renders their classification more difficult.

\paragraph{\bf (HD\,87643)}
The near-IR colors of HD\,87643 place it to the high end of the HAeBe regime. However, these colors have 
large errors and require refinement. Studies based on long-baseline interferometry revealed that this 
object consists of two B-type stars, and each component might be surrounded by a dusty disk 
(\cite{2009A&A...507..317M}). Whether this object is a physical binary, is currently not known.

\paragraph{\bf (HD\,62623)}
It appeared in the literature as the first A[e]SG (\cite{2010A&A...512A..73M, 2011A&A...526A.107M}) and 
is known to be surrounded by a detached gas and dust disk, from which also CO band emission has been 
detected (\cite{2012BAAA...55..123M, 2018MNRAS.480..320M}). It has been speculated that the gap between 
the star and the inner rim of the gas disk might have been cleared by a companion. But no clear evidence 
of such a companion has been found yet. Its low luminosity and large errors in the near-IR photometric 
measurements make it challenging to unambiguously assign the star a B[e]SG status, although its position 
within the near-IR seems to coincide more with the B[e]SG domain rather than with any other 
classification. This object is an ideal candidate to clarify its nature based on dedicated $K$-band 
observations to analysis the $^{13}$CO content within its circumstellar disk.


\section{Discussion and Conclusions}\label{sect:discussion}

B[e]SGs form a special class of evolved massive stars and are thought to represent a short-lived 
transition phase either in their red-ward, post-main sequence evolution, or in their blue-ward post-RSG 
evolution. The total number of currently known B[e]SGs is low, supporting the idea of a short 
transition phase. 
But in which direction the stars evolve, and whether all B[e]SGs evolve into the same 
direction, is still an open issue. For the objects in M31 and M33 it has been argued that the B[e]SGs are 
more isolated than LBVs and hardly found in stellar associations, so that a post-RSG (or post-yellow 
supergiant) evolution was proposed to be more likely (\cite{2017ApJ...844...40H}). This assessment was based 
on the available numbers of putative B[e]SGs in these two galaxies at that time. But after revision of the 
two samples (Tables\,\ref{tab:M31} and \ref{tab:M33}), two of the four confirmed B[e]SGs in M31 
and both confirmed B[e]SGs in M33 are associated with stellar groups, questioning the conclusion that 
B[e]SGs are isolated and thus post-RSGs. 

The best age indicator for B[e]SGs we have to date is their surface abundance enrichment in $^{13}$C, as 
discussed in Section \ref{sect:current_evol}. If B[e]SGs would be post-RSGs, their progenitors would 
all have started with a (very) low rotation speed. While such a scenario cannot be excluded, it may not be
very likely considering that stars are born on average with a rotation rate of about 40\% of their 
critical velocity (\cite{2012A&A...537A.146E}).

Interaction within a (close) binary system, maybe even up to a binary merger, seems to 
be an alternative and popular scenario (e.g., \cite{2012A&A...545L..10W, 2014AdAst2014E..10D}), but the 
number of currently confirmed binaries amongst B[e]SGs is still rather low to give preference to the 
binary channel as the sole possible way for the formation of B[e]SGs. Likewise, some B[e]SGs have been 
suggested as suitable supernova candidates. For example, the Galactic object MWC\,137 appears similar to 
Sher 25 (\cite{2008MNRAS.388.1127H}) and SBW1 (\cite{2013MNRAS.429.1324S}), which both look like the 
progenitor of SN1987A. Also the SMC object LHA\,115-S\,18 has been proposed to be a viable SN1987A 
progenitor (\cite{2013A&A...560A..10C}). In this respect, it is vital to resolve B[e]SG populations and to 
study their properties.

In this review, a census of the currently known B[e]SG population in the Milky Way and in nearby 
star-forming galaxies within and beyond the Local Group is presented. The proposed candidates have been 
undertaken a critical examination, sorted into confirmed B[e]SGs and B[e]SG candidates, and unsuitable 
objects have been flagged as "misclassified"\footnote{We would like to caution that with insufficient
knowledge of stellar properties, individual objects may easily be misclassified, as it happened 
in recently published catalogs (\cite{2015MNRAS.447..598S, 2018AJ....156..294A, 2016AJ....152...62M, 
2018RNAAS...2c.121R}), in which erroneously a number of (even confirmed) B[e]SGs are listed as LBV
candidates.}. 

During these investigations, a fundamental difference has been recognized between the identification 
issues for objects in our Galaxy compared to those in other galaxies. Extragalactic B[e]SGs bare the 
risk of being confused with LBVs in quiescence, which share very similar optical spectroscopic characteristics. To 
separate these two classes of objects, one can make use of clearly defined classification criteria based 
on certain sets of emission features identified in their optical and near-IR spectra as was outlined in 
Section\,\ref{sect:method}. In addition to these, inspection of the location of possible B[e]SG 
candidates in the IR color-color diagrams is highly advisable, because B[e]SGs and LBVs populate 
clearly separate domains. This fact might also be used as starting point for future investigations
of extragalactic samples. However, for such future studies, infrared photometry
with higher spatial resolution than what is currently provided by 2MASS and WISE is desirable to prevent
from contamination with neighboring sources in densely populated regions. Moreover, precise distances, as 
soon provided by GAIA, will help to separate foreground stars in the directions to other galaxies that 
might have been misclassified as luminous extragalactic stars.

In the Milky Way, an additional complication occurs due to the often uncertain luminosity estimates, and 
the overlap of low-luminosity B[e]SGs with the most luminous massive pre-main sequence objects (HAeBe). 
Here, special care needs to be taken, especially since both classes of objects occupy adjacent regions
in the near-IR color-color diagram with a probable overlap. Without additional distinctive features for
such low-luminosity, borderline B[e]SG candidates, their real nature remains elusive. One complementary 
classification criteria that was discussed, is provided by the enrichment of the circumstellar material 
of evolved objects with processed material that has been released from the stellar surface, as opposed 
to the non-processed material with interstellar abundance patterns found around HAeBes. The most ideal 
element to search for is $^{13}$C, which is bound in $^{13}$CO molecules in the circumstellar disks. 
Measuring the $^{13}$CO amount with respect to $^{12}$CO provides immediate insight into the nature of 
the object. The current HAeBe samples might hide such low-luminosity 
B[e]SG candidates, which can only be identified as such by careful and honest analysis.

The result from this census, after strict application of the classification criteria, is that we count 
nine confirmed and six candidate B[e]SGs in the Galaxy. Moving out to the MCs, the numbers amount to 13 
(+2) in the LMC and 5 (+1) in the SMC. The situation in other members of the Local Group is not much 
better. There the numbers drop to 4 (+7) in M31 and 2 (+6) in M33. Even further away, only 3 candidates 
have been reported from M81. The total number of B[e]SGs found in the various galaxies are too small for 
statistical analyses with respect to a metallicity dependence of their number, but the comparable 
quantities within the Milky Way and the LMC suggest only a mild dependence of the amount of B[e]SGs on 
(i.e. a possible drop with decreasing) metallicity. 

Further, we report that the Galactic sample of 15 B[e]SGs and candidates contains currently seven 
confirmed or suspected binaries (see Table\,\ref{tab:Galactic}), which is less than half of the 
population. In the MCs this number is even lower, because so far only three stars have been reported to 
be possible binaries, of which one, the SMC star LHA\,115-S\,6 (= RMC\,4) has been proposed to be the 
remnant of a binary merger within an initially triple system (\cite{1998ASSL..233..235L, 
2000AJ....119.1352P}). The other two objects, the LMC star LHA\,120-S\,134 and the SMC object 
LHA\,115-S\,18, have been identified as optical counterparts of X-ray sources (\cite{2013A&A...560A..10C, 
2015salt.confE..55B}). Nothing is known about possible binarity in the B[e]SG samples from the other 
galaxies, although photometric variability was seen in at least two M31 objects: the confirmed B[e]SG star 
J004417.10+411928.0 (\cite{2015MNRAS.447.2459S, 2017AJ....154...81M}) and the candidate 
J004444.52+412804.0 (\cite{1999AJ....117.2810S, 2006A&A...459..321V}). Whether this variability is a sign 
of binarity or just of semi-regular variability which might be interpreted with pulsation activity such as 
reported from the $\alpha$\,Cygni variables, needs to be studied in more detail. 
 
For the sake of completeness, a special class of objects, which have not been discussed yet, should be 
briefly mentioned as well. These are the high-mass X-ray binaries (HMXB) with possible B[e] supergiant 
(candidate) companion. Some objects with an assigned B[e]SG status have been found to be too luminous in 
X-rays for being considered single stars. These objects have been proposed to be binary systems, in 
which the high energy emission is caused by either accretion onto a compact object, or by shocks in a 
colliding wind binary with a second massive star. Members of this group are the Galactic objects 
Cl*\,Westerlund\,1\,W\,9 (= Wd1-9), which is considered a colliding wind system 
(\cite{2013A&A...560A..11C}), CI\,Cam (= MWC\,84), which might be interpreted as supernova imposter 
(\cite{2019A&A...622A..93B}), and the high-mass X-ray binary (HMXB) IGR J16318-4848 in which the compact 
object was proposed to be a neutron star (\cite{2004ApJ...616..469F}). Two ultra-luminous X-ray sources 
(ULXs), Holmberg II X-1 and NGC 300 ULX1, the latter being also named as supernova imposter SN2010da, have 
also been proposed to be HMXBs including a B[e]SG (\cite{2016ApJ...830..142L, 2017ApJ...838L..17L, 
2016ApJ...830...11V}). Whether these objects host indeed a B[e]SG clearly needs to be 
investigated in more detail. But since the behavior of these sources is considerably different from the 
confirmed B[e]SGs, I hesitate to include them into the census.

The presented populations represent the current knowledge of B[e]SGs and B[e]SG candidates in the 
closest star-forming galaxies. With the ever growing sensitivities of instruments and telescope sizes, 
the future in B[e]SG star research is bright, because many more candidates will be identified in even 
more distant galaxies and with metallicities spreading over a large range. With statistically meaningful
samples it will finally be possible to unveil the nature and fate of these fascinating objects.

\vspace{6pt} 




\funding{This project has received funding from the Grant Agency of the Czech Republic (GA\v{C}R, grant number 17-02337S). The Astronomical Institute of the Czech Academy of Sciences, Ond\v{r}ejov, is supported by the project RVO:67985815.}

\acknowledgments{I wish to thank the editor, Roberta Humphreys, for the invitation to write this review.
In addition, I am grateful to Michalis Kourniotis for inspiring discussions on searching for  
extragalactic B[e]SGs, and to Dieter Nickeler and Lydia Cidale for passionate discussions about 
B[e]SGs as well as for  their proof-reading of and suggestions on the draft versions. Moreover, I thank 
the anonymous referees for their careful reading and suggestions on the draft version. 
This research made use of the NASA 
Astrophysics Data System (ADS) and of the SIMBAD 
database, operated at CDS, Strasbourg, France. 
This publication makes use of data products from the 2MASS, which
is a joint project of the University of Massachusetts and the
Infrared Processing and Analysis Center/California Institute
of Technology, funded by the National Aeronautics and Space
Administration and the National Science Foundation.
This publication makes use of data products from the Wide-field 
Infrared Survey Explorer, which is a joint project of the 
University of California, Los Angeles, and the Jet Propulsion 
Laboratory/California Institute of Technology, funded by the 
National Aeronautics and Space Administration. }

\conflictsofinterest{The author declares no conflict of interest.} 

%


\reftitle{References}






\end{document}